\documentclass[default,iicol]{sn-jnl}
\usepackage{graphicx}%
\usepackage{multirow}%
\usepackage{amsmath,amssymb,amsfonts}%
\usepackage{amsthm}%
\usepackage{mathrsfs}%
\usepackage[title]{appendix}%
\usepackage{xcolor}%
\usepackage{textcomp}%
\usepackage{manyfoot}%
\usepackage{booktabs}%
\usepackage{algorithm}%
\usepackage{algorithmicx}%
\usepackage{algpseudocode}%
\usepackage{listings}%
\usepackage{tocloft}
\usepackage{float}
\usepackage{upgreek}
\usepackage[export]{adjustbox}

\theoremstyle{thmstyleone}%

\theoremstyle{thmstyletwo}%

\theoremstyle{thmstylethree}%

\usepackage{xcolor}
\definecolor{bleu}{rgb}{0, 0, 0}
\title[ ]{Control of collective activity to crystallize an oscillator gas}
\begin{document}

\author[1]{\fnm{Marine} \sur{Le Blay}}\email{leblay@physics.leidenuniv.nl}
\author[1]{\fnm{Joshua H. K.} \sur{Saldi}}\email{saldi@physics.leidenuniv.nl}
\author*[1]{\fnm{Alexandre} \sur{Morin}}\email{morin@physics.leidenuniv.nl}

\affil[1]{\orgdiv{Soft Matter Physics, Huygens-Kamerlingh Onnes Laboratory}, \orgname{Leiden University}, \orgaddress{\street{P.O. Box 9504}, \city{Leiden}, \postcode{2300 RA}, \country{The Netherlands}}}

\abstract{
Motility-induced phase separation occurs in assemblies of self-propelled units when activity is coupled negatively to density. In contrast, the consequences of a positive coupling between density and activity on the collective behaviour of active matter remain unexplored. Here, we show that collective activity can emerge from such a positive coupling among non-motile building blocks. We perform experiments with self-sustained oscillators powered by contact-charge electrophoresis. Although the oscillators are non-motile by design, they spontaneously form an active gas when confined together. The super-elastic nature of collisions constitutes a positive density-activity coupling and underlies the active gas properties. Elucidating the origin of binary collisions allows us to precisely control the structure of the active gas and its eventual crystallization. Beyond considering the overlooked positive coupling between density and activity, our work suggests that rich collective properties can emerge not only from the symmetry of interactions between active building blocks, but also from their adaptable and responsive behaviour.
}


\maketitle

\clearpage
\begin{figure*}[h]
    \centering
    \includegraphics[width=\textwidth]{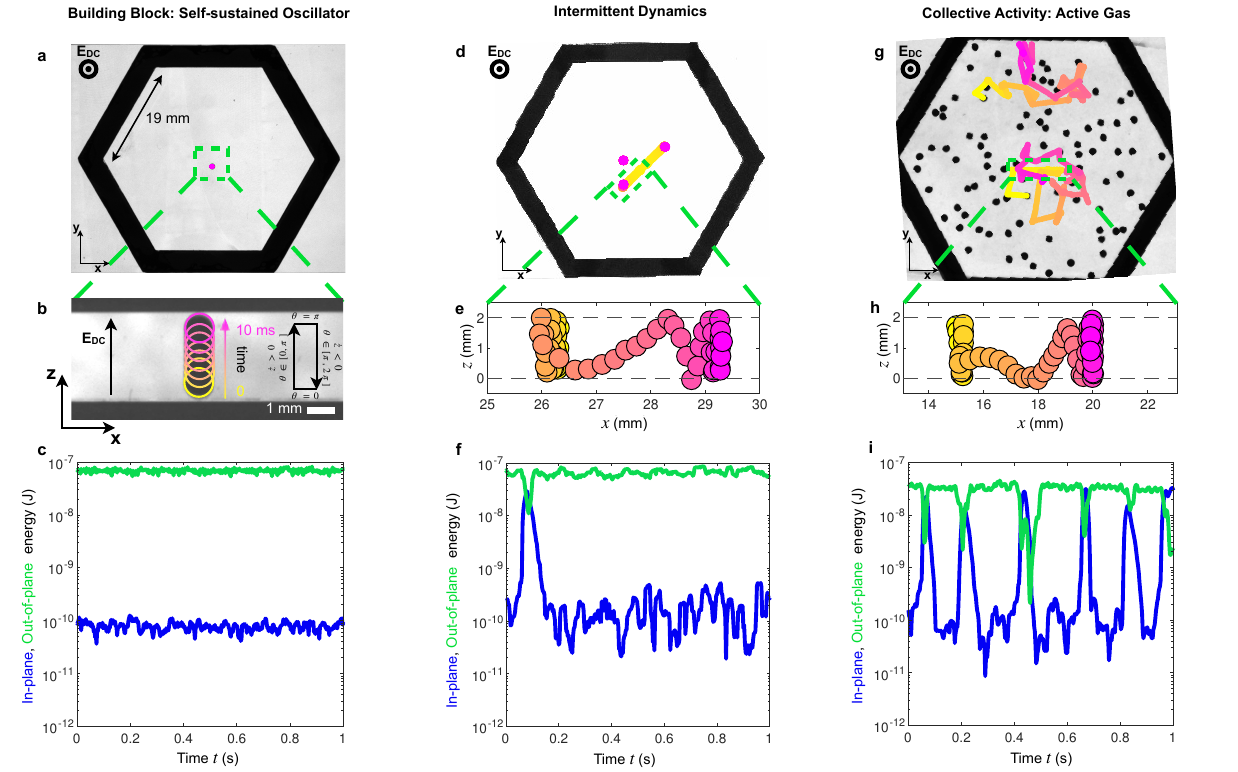}
    \caption{
    {\bf Collective activity: from a steady self-sustained oscillator to an active gas.  $\lvert$ }
			{\bf (a-c) Single self-sustained oscillator.} A $1\,\rm mm$-diameter sphere is turned into a self-sustained oscillator through the application of an electric field $\mathbf{E_{\rm DC}}$ via contact-charge electrophoresis. It is confined within an hexagonal boundary (black).  
            {\bf (a)} Top view, along the electric field direction. The oscillator is steady in space, as revealed by the $3\,\rm s$-long trajectory displayed.
            {\bf (b)} Side view, perpendicular to $\mathbf{E_{\rm DC}}$: the sphere oscillates back and forth between the two electrodes with a period of $\sim 20\,\rm ms$. \textcolor{bleu}{The phase $\theta$ naturally parametrizes its oscillatory motion, see Methods.}
            {\bf (c)} The \textcolor{bleu}{out-of-plane kinetic energy $K_\perp$} is several orders of magnitudes larger than the \textcolor{bleu}{in-plane kinetic energy $K_\parallel$}. The dynamic is virtually noiseless, both energies barely fluctuate in time.
			{\bf (d-f) Intermittent dynamics.}
			{\bf (d)} When few oscillators - here three - are confined together, they display an intermittent dynamics. $1\,\rm s$-trajectories are shown (from yellow to pink).
			{\bf (e)} Stereo-vision reconstructed trajectory of the oscillator highlighted in (d). $150\,\rm ms$-trajectories are shown (from yellow to pink)
			{\bf (f)} The time evolution of \textcolor{bleu}{$K_\perp$ and $K_\parallel$} captures the intermittent dynamics of oscillators.
			{\bf(g-i) Active oscillator gas.}
			{\bf (g)} Snapshot of the active oscillator gas at $\phi = 9\,\rm \%$ with two trajectories revealing the fast dynamics of the gas (color lines, $2\rm\,s$ from yellow to pink).
			{\bf (h)} Projection in the (xz)-plane of the trajectory of the oscillator highlighted in (g) obtained by stereo-vision ($150\rm\,ms$ from yellow to pink). The oscillator alternates between oscillating in place and performing large displacements.
			{\bf (i)} The \textcolor{bleu}{out-of-plane kinetic energy $K_\perp$} is transferred repeatedly to \textcolor{bleu}{in-plane kinetic energy $K_\parallel$, which fuels mobility. $E_{\rm DC} = 0.87\,\rm kV\,mm^{-1}$ in all panels.}.
    }
    \label{Figure:1}
\end{figure*}
\textcolor{bleu}{If mobility is one of the hallmarks of living organisms, their efficient use of resources and energy is undoubtedly widespread and equally vital.
For example, a prey is better off saving some energy to be able to escape from predators, rather than exhausting itself in unnecessary and frantic wanderings.
Yet, at odds with any notion of frugality, active matter physicists have focused on studying collectives of active constituents with constant levels of activity.
Under these conditions, differences in interaction rules between constituents lead to a wealth of dynamical states, such as flocking~\cite{toner2005hydrodynamics,bricard2013emergence} and active turbulence~\cite{doostmohammadi2018active,duclos2020topological,alert2022active}, whose broken symmetries reflect the symmetry of interactions~\cite{deseigne2010collective,suzuki2015polar,yan2016reconfiguring,najma2024microscopic,maitra2020chiral,aubret2021metamachines,baconnier2022selective,giomi2022long,armengol2023epithelia}.
Despite the success of this approach, it cannot be comprehensive, as it fails to represent self-organization originating from changes in activity levels upon interactions.
Discovering the emergent properties of condensed active matter with adaptable and responsive activity remains, however, a formidable challenge~\cite{mijalkov2016engineering,lavergne2019group,alston2022intermittent,van2023self,ketzetzi2024self}.}

\textcolor{bleu}{Here, we tackle this challenge and discover the phenomenon of collective activity, i.e. the emergence of mobility within populations of individually non-motile constituents.}

\textcolor{bleu}{Our discovery stems from the study of a model active material assembled from self-oscillating particles.
We introduce the experimental system in the first part of this Article and explain how their oscillation kinetic energy acts like an energy reserve, which is only released on interactions, powering their mobility.
Interestingly, this responsive activity behaviour opposes the behaviour of some bacteria with quorum-sensing ability: sensing high density in their surroundings, they reduce their mobility which promotes biofilm formation~\cite{parsek2005sociomicrobiology,guttenplan2013regulation,grobas2021swarming,worlitzer2022biophysical}.
While the well-studied framework of motility-induced phase separation~\cite{farrell2012pattern,theurkauff2012dynamic,buttinoni2013dynamical,cates2015motility,bauerle2018self,liu2019self} captures such a {\em negative} coupling between density and mobility, the reverse scenario of a {\em positive} coupling remains overlooked.
By sharing our findings on collections of self-sustained oscillators with collective activity, we launch the study of this new class of active materials where density enhances individual activity.}
\textcolor{bleu}{
Here, we further explore both the emergent properties and the control opportunity arising from collective activity:
we rationalize the emergence of an active gas with athermal statistics, and tame interaction-induced mobility to guide its crystallization.}

\textcolor{bleu}{
Overall our work illustrates how responsive active behaviours lead to novel self-organization pathways which can be harnessed to form new states of active matter.
}

\section*{Collective activity}
\textcolor{bleu}{We now introduce} our experimental system featuring collective activity, illustrated in Fig.~\ref{Figure:1}.
{\textcolor{bleu}{Unlike in typical active systems,} our building-blocks are \textcolor{bleu}{non-motile} particles: \textcolor{bleu}{they consist in} self-sustained oscillators whose internal dynamics is parametrized by their phase $\theta$, see Fig.~\ref{Figure:1}(a-c).
Experimentally, \textcolor{bleu}{we turn millimetric metal spheres into oscillators by taking advantage of the so-called contact-charge electrophoresis effect~\cite{tobazeon1996electrohydrodynamic,mersch2011antiphase,drews2015contact,eslami2016modeling,dou2018emergence}:
the motion of particles, which having acquired a charge in contact with an electrode, are driven by an electric field.
Specifically, we confine stainless steel spherical beads between two electrodes.
The beads have a diameter $2R = 1\,\rm mm$, the electrodes are spaced by $h=3\,\rm mm$, the device is filled with oil, and we apply a constant electric field $\mathbf{E_{\rm DC}}$ (see Methods).
Above a threshold value $E^\star = 0.72\,\rm kV\,mm^{-1}$, the electric force overcomes gravity and the particles start oscillating between the two electrodes, see Fig.~\ref{Figure:1}(b) and Supplementary Video 1.
After a short transient, a state of steady oscillations is reached, with a well-defined oscillation frequency whose typical value is $ \omega_{\rm o} = 400\,\rm rad\,s^{-1} $ under $\lvert\mathbf{E_{\rm DC}}\rvert = 10^6\,\rm V\,m^{-1} $.
This oscillatory motion is naturally parametrized by a phase $\theta \in [0,\,2\pi[$, see Fig.~\ref{Figure:1}(b) and Methods.
We further report on the driving mechanism in \textcolor{bleu}{Supplementary Information (SI)}, and on the transient dynamics and the variation of the oscillation frequency with the electric field in Extended Data~Fig.~\ref{Figure:ExtData_SingleOscilator}.}

\textcolor{bleu}{
The possibility to tune the oscillation frequency is particularly beneficial to our purpose.
Considering that the kinetic energy of oscillation amounts to an energy storage, this setup gives us a convenient mean to change the capacity of this energy reserve.
At the same time, we stress that individual oscillators are non-motile and barely fluctuate in the plane orthogonal to their oscillation direction. 
As shown in Fig.~\ref{Figure:1}(c), their in-plane kinetic energy $K_\parallel$ is several order of magnitudes lower than their out-of-plane kinetic energy of oscillation $K_\perp$ (see Methods).
}

While individual oscillators remain spatially put, they get mobile when confined together, a behaviour that we term collective activity, see Supplementary Video 1.
Indeed, collectives of only a few oscillators are already mobile, but the resulting dynamic is highly intermittent, as illustrated in Fig.~\ref{Figure:1}(d-f). 
Collective activity hence really kicks in at higher density, as shown in Fig.~\ref{Figure:1}(g-i).
Oscillators confined at a packing fraction $\phi = 9\,\%$ quickly reach a dynamical steady-state and spontaneously form an active gas featuring large displacements of the building-blocks and a highly disordered structure.
Within this active gas, we follow the three-dimensional trajectories of all oscillators by means of stereo-imaging (see Methods).
The time evolution of \textcolor{bleu}{$K_\perp$ and $K_\parallel$ reveals that in-plane mobility is} activated by a transfer of energy from the \textcolor{bleu}{out-of-plane degree of freedom 
($\theta$) to the in-plane ones (${x,y}$)}, see Fig.~\ref{Figure:1}(i).
Interestingly, this transfer echoes the theoretical work by Han et al.~\cite{han2021fluctuating} that examines the coupling between ``activated''  and ``fluctuating'' degrees of freedom and the resulting equilibrium-like properties.
Unlike their findings, however, the emerging properties of the active oscillator gas are athermal.
Indeed, we next unravel the connections between the energy flow, \textcolor{bleu}{from out-of-plane to in-plane degrees of freedom}, and the non-equilibrium statistical properties of the active gas.

\section*{Active gas fuelled by super-elastic collisions}
\begin{figure*}[h]
    \centering
    \includegraphics[width=\textwidth]{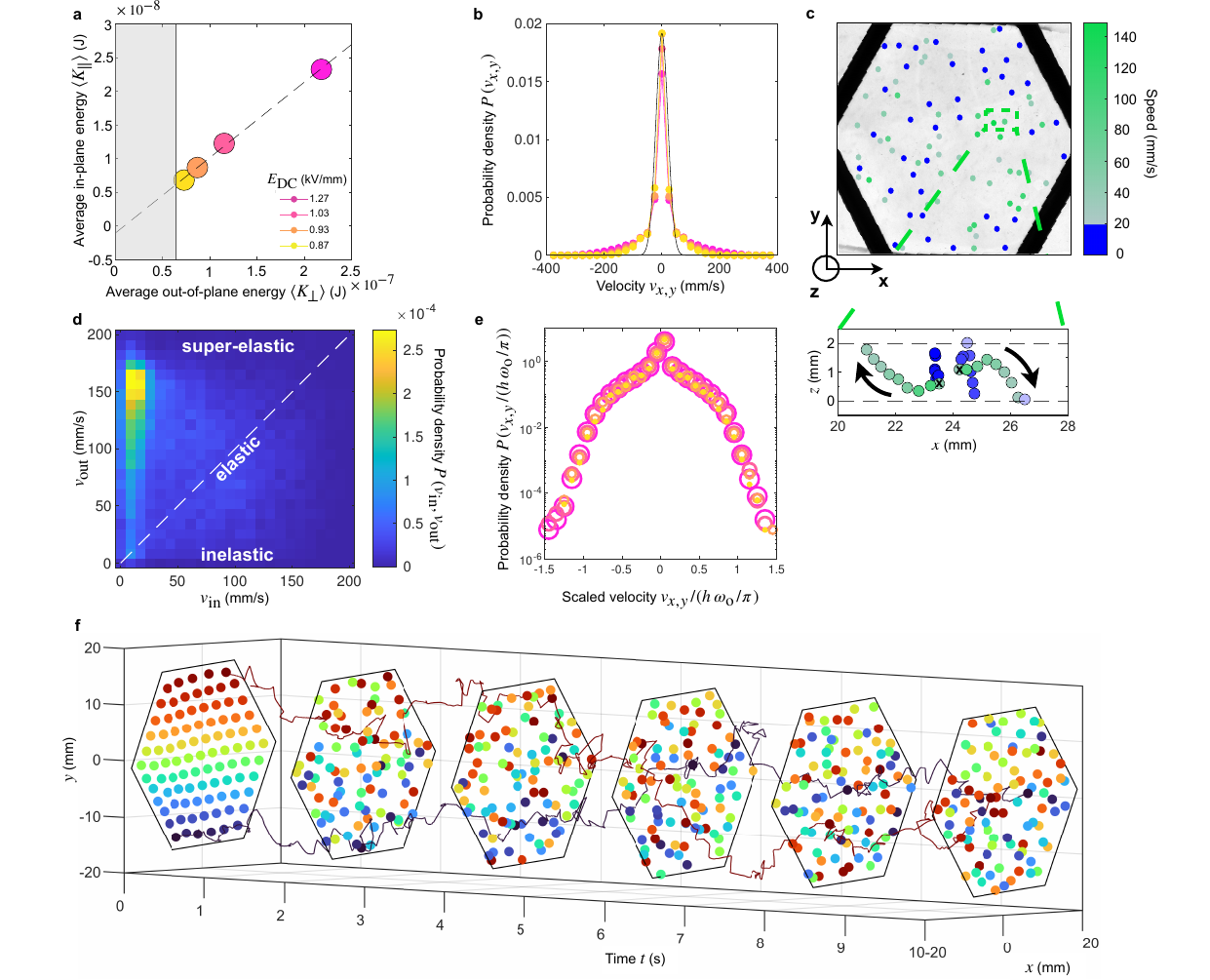}
    \caption{
    {\bf Active gas properties follow from super-elastic collisions. $\lvert$}
			{\bf (a)} The total \textcolor{bleu}{in-plane kinetic energy} of the active gas scales linearly with the \textcolor{bleu}{out-of-plane kinetic energy}, varied through the electric field amplitude $\lvert\mathbf{E_{\rm DC}}\lvert$.  \textcolor{bleu}{The grey area is not accessible since oscillations only occur above $E^\star$. The dotted line is an affine fit to the data points.}
			{\bf (b)} Probability density functions (PDFs) of the velocity components $v_x$ and $v_y$. The color codes for the out-of-plane kinetic energy as in (a). Black line: normal distribution. The PDFs are non-Gaussian, with fat tails, and confirm the non-equilibrium nature of the active gas.
			{\bf (c)} Slow-moving oscillators (\textcolor{bleu}{blue}) coexist with fast-moving ones (\textcolor{bleu}{green}). Inset: the projection in the (xz)-planes of the trajectories of two fast-moving particles unveils their past collision, marked with crosses. The arrows indicate the directions of motion after the collision. $\lvert\mathbf{E_{\rm DC}}\rvert = 0.87\rm\,kV\,mm^{-1}$.
			{\bf (d)} Super-elastic collisions are the most frequent type of binary collisions, as evidenced by the joint probability density $P(v_{\rm in},v_{\rm out})$. $\lvert\mathbf{E_{\rm DC}}\rvert = 0.87\rm\,kV\,mm^{-1}$.
			{\bf (e)} All probability density functions collapse when scaling the velocity by the oscillation frequency. This non-trivial collapse takes its origin in the scaling with $\omega_{\rm o}$ of both energy injection, through super-elastic collisions, and energy dissipation. The color codes for the \textcolor{bleu}{out-of-plane kinetic energy} as in (a) and (b).
			{\bf (f)} Collective activity quickly destroys an initially ordered crystal. Worldlines of two oscillators are drawn and connect six subsequent snapshots. The different colors identify each oscillator.
    }
    \label{Figure:2}
\end{figure*}
To investigate the properties of the active gas, we vary the oscillation frequency $\omega_{\rm o}$ via the amplitude of the electric field, at constant density $\phi = 9\,\%$.
Figure~\ref{Figure:2}(a) shows that the \textcolor{bleu}{total in-plane kinetic energy} increases linearly with the \textcolor{bleu}{total out-of-plane kinetic energy, confirming that  $xy$-displacements} are fuelled by the intrinsic dynamics of the building blocks \textcolor{bleu}{which acts like an energy reserve}.
Accordingly, the distribution of velocities $v_{x,y}$ shown in Fig.~\ref{Figure:2}(b) broadens with $\omega_{\rm o}$.
We stress, however, that this behaviour does {\em not} result from an increase in (effective) temperature: i) the individual dynamics is virtually noiseless (Fig.~\ref{Figure:1}(a)-(c) and SI), ii) the dynamics is not time-reversal symmetric (Supplementary Video 2), and iii) the velocity distributions depart from Maxwell-Boltzmann statistics (Fig.~\ref{Figure:2}(b)).

The shape of the distributions, with fat tails, hints at a peculiar dynamics with broadly-distributed fast-moving particles and many slow-moving ones.
The distribution of speeds among particles shown in Fig.~\ref{Figure:2}(c) reveals that the two populations coexist in space, a visually striking feature of the active gas, see Supplementary Video 2.
Looking back in time at the trajectories of mobile particles shows that they were initially oscillating in place, before colliding with one another and acquiring \textcolor{bleu}{in-plane} velocity.
This observation suggests that binary collisions are instrumental in \textcolor{bleu}{collective activity}. 
We therefore proceed with studying binary collisions to capture the statistical properties of the active gas.
Analysing more than $30000$ collisions, we build in Fig.~\ref{Figure:2}(d) the joint probability density $P(v_{\rm in},v_{\rm out})$ of an oscillator exiting a collision with a transverse speed $v_{\rm out}$ given that it entered with a speed $v_{\rm in}$.
It reveals that generally $v_{\rm out} > v_{\rm in}$, {\it i.e.} most collisions are effectively super-elastic.

The dynamics of the active gas is therefore fuelled by collisions.
In steady-state, the energy that these collisions inject must be balanced by dissipation processes.
We demonstrate in the \textcolor{bleu}{SI that energy injection due to super-elastic collisions between particles and energy dissipation due to inelastic collisions with the electrodes} both scale with the oscillation frequency $\omega_{\rm o}$.
\textcolor{bleu}{Even more so, taken together, these scalings imply the rescaling of the velocity distribution with $\omega_{\rm o}$.
A behaviour we confirm experimentally} in Fig.~\ref{Figure:2}(e):
Increasing the oscillator frequency merely accelerate the collective dynamics, but does not change its nature.

Altogether our findings point out the specificity of the active gas:
unlike thermal systems and active systems with active forces, mobility \textcolor{bleu}{is here an emergent property.}
\color{bleu}
This emergent dynamics is the result of a strong and positive coupling between density and activity induced by
super-elastic collisions that release the energy stored by the constituents.
This positive coupling is particularly efficient in imposing disorder: even when preparing the system in a crystalline state, order is quickly lost (see Fig.~\ref{Figure:2}(f)).
Because of its importance on the collective dynamics, it constitutes a prominent control opportunity at the same time.
We explore this opportunity in the remainder of this Article, and ask:
Could we tame collective activity to gain control over the emerging dynamics?
}

\color{black}
\section*{Synchronization-dependent interactions}

\begin{figure*}[ht!]
    \centering
    \includegraphics[width=\textwidth]{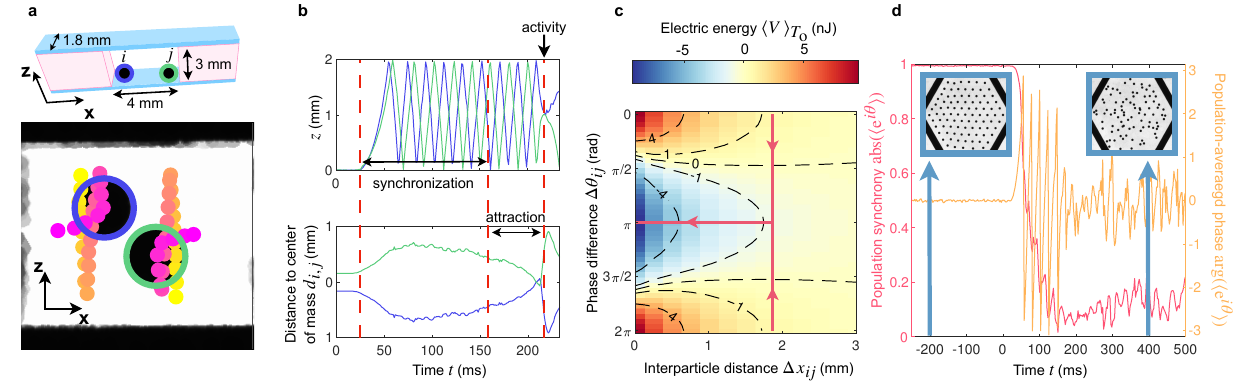}
    \caption{
{\bf Collisions origin from synchronization-dependent interactions between oscillators. $\lvert$}
			{\bf (a)} Top: perspective view of the quasi-1D channel used to confined two oscillators. Bottom: experimental snapshot of the two oscillators (circled in blue and green) with their trajectories overlaid ($30\,\rm ms$, from yellow to pink ).
			{\bf (b)} Two-stage dynamics of the oscillator pair: synchronization and attraction. Top: the time evolution of the oscillator heights shows that the oscillators synchronize in phase opposition in $\sim 150\,\rm ms$.  Bottom: the time evolution of the oscillator distance to the \textcolor{bleu}{centre of mass $d_{\rm i,j} = x_{\rm i,j}-(x_i +x_j)/2-R$} shows that the oscillators subsequently attract each other. At $t \simeq 220\,\rm ms$, the oscillators collide and move away from each other. Blue: left oscillator. Green: right oscillator.
			{\bf (c)} Electric potential energy of the oscillator pair. The electric energy is obtained by finite element simulations for various interparticle distances \textcolor{bleu}{$\Delta x_{ij} = \lvert x_j-x_i \rvert - 2R$} and phase differences \textcolor{bleu}{$\Delta \theta_{ij}=\theta_j-\theta_i$}, and is averaged over one oscillation period\textcolor{bleu}{, see Methods}.
            When in phase \textcolor{bleu}{($\Delta\theta_{ij} = 0,\,2\pi$)}, electric interaction causes the oscillators to spatially repel. When in phase opposition \textcolor{bleu}{($\Delta\theta_{ij} = \pi$)}, electric interaction causes the oscillators to attract.
            Red line: most probable experimental path.
			{\bf (d)} Transient evolution of the collective state ($\phi = 9\,\rm \%$). The electric field $\mathbf{E_{\rm DC}}$ is turned on at $t=0$. Red, left axis: norm of the synchronization order parameter \textcolor{bleu}{$\lvert \mathbf{S} \rvert = {\rm abs}(<{\rm exp}({i\theta_i)}>_i)$, quantifying population synchrony}. Orange, right axis: phase of the synchronization order parameter \textcolor{bleu}{${\rm arg}(\mathbf{S}) = {\rm arg}(<{\rm exp}({i\theta_i)}>_i)$}. The loss of global synchrony occurs in $\sim 200\,\rm ms$. Insets: snapshots at $t=-200\,\rm ms$ and $t=400\,\rm ms$.
    }
    \label{Figure:3}
\end{figure*}

To answer \textcolor{bleu}{this question}, the mechanism leading to collisions between initially static oscillators must be uncovered.
\textcolor{bleu}{We perform experiments on a pair of particles confined together to access the detailed dynamics of their interactions.
As shown in Fig.~\ref{Figure:3}(a), their in-plane dynamics is restricted to the $x$-axis.}
As the electric field is applied and the particles start oscillating, we observe a two-step dynamics, see Fig.~\ref{Figure:3}(b).
First, synchronization takes place, with the particles gradually going in phase opposition.
Second, the two particles come closer until they collide.
Since particles switch charge polarity when reaching an electrode, particles in phase opposition bear opposite charges, which causes attraction.
We confirm this mechanism in Fig.~\ref{Figure:3}(c), which shows the electric potential energy obtained by finite element simulation and integrated over the fast oscillation dynamics \textcolor{bleu}{(see Methods)}.
Interestingly, it reveals a synchronization-dependent interaction with particles repelling for small phase-shifts $\lvert \delta\theta \lvert \lesssim
 \pi/2$ and attracting for larger ones $\lvert \delta\theta \lvert \gtrsim \pi/2$, which matches experimental measurements \textcolor{bleu}{(see also Extended Data Fig.~\ref{Figure:SI_2})}.
 \textcolor{bleu}{We unveil this interplay between phase and space dynamics by establishing the equations of motion for the phase $\theta_i$ and position $x_i$ of particle $i$ from first principles (see SI).
 Keeping only lower order contributions and coarse-graining over one oscillation period, the equations of motion reduce to a Kuramoto evolution of $\theta_i$ coupled to an underdamped dynamics of $x_i$. 
 They read:
 }
 \begin{align}
     & \frac{\partial^2 \theta_i}{\partial t^2} = \frac{\omega_{\rm o}}{T_{\theta}} - \frac{1}{T_\theta}\frac{\partial \theta_i}{\partial t} + K_{ij} \sin\left( \theta_j - \theta_i \right), \label{Eq:Phase} \\
     & \frac{\partial^2 x_i}{\partial t^2} = - \frac{1}{T_x}\frac{\partial x_i}{\partial t} + J_{ij} \cos\left( \theta_j - \theta_i \right), \label{Eq:Space}
 \end{align}
where $T_\theta$ and $T_x$ are the characteristic time scales of dissipation along the \textcolor{bleu}{out-of-plane and in-plane directions, respectively.}
\textcolor{bleu}{
The electric interactions between oscillators $i$ and $j$ are captured by the coupling terms $K_{ij}(\Delta x_{ij})=K_0 e^{- \Delta x_{ij}/\lambda}$ , and $J_{ij}(\Delta x_{ij}) = -\frac{h-2R}{\pi}\frac{\partial K_{ij}}{\partial x_i}$, with $\Delta x_{ij} = \lvert x_{j}-x_{i} \rvert - 2R$ (see SI).
We determine $K_0 = -4\,\rm{rad\,s}^{-2}$ and $\lambda=8\times 10^{-4}\,\rm{m}$ from finite element simulations (see Fig.~\ref{Figure:3}(c) and SI).
Since $K_{ij}<0$, Eq.~(\ref{Eq:Phase}) promotes phase opposition between oscillators, while the cosine dependency in Eq.~(\ref{Eq:Space}) captures a switch from repulsion to attraction as the phase difference increases.}
Interestingly, the bi-directional coupling of Eqs.~(\ref{Eq:Phase}-\ref{Eq:Space}) is reminiscent of the so-called \textcolor{bleu}{``swarmalator''~\cite{o2017oscillators} and ``pulsating active matter'' models~\cite{zhang2023pulsating}}.
For pairs of oscillators, however, the switching from repulsive to attractive interactions is unparalleled.
As it eventually yields to super-elastic collisions, the dynamics of Eqs.~(\ref{Eq:Phase}-\ref{Eq:Space}) and comparatively weak electric interactions are only valid before collisions occur and dominate the collective dynamics.

\textcolor{bleu}{The binary dynamics Eqs.~(\ref{Eq:Phase}-\ref{Eq:Space}) is highly relevant when it comes to the initial stages of the collective dynamics ($\phi = 9\,\%$).}
\textcolor{bleu}{Firstly, its onset consists in pairs of oscillators slowly attracting each others and colliding with opposite phases, see Extended Data Fig.~\ref{Figure:ED_Onset} and Supplementary Video~3.
Secondly, the evolution of the norm and phase of the synchronization order parameter $\mathbf{S} = \langle \exp{ \left(i\theta_i \right) } \rangle_i$, shown in Fig.~\ref{Figure:3}(d), reveals a quick lose of synchrony.
While particles start oscillating in phase, global phase incoherence is reached in about $200\,\rm ms$, a duration commensurate with the binary collision time (see Fig.~\ref{Figure:3}(b)).}
It is therefore the interplay between phase and space dynamics resulting from electric interactions that steers the oscillator collective towards forming an active gas.
The synchronization-dependent nature of interaction between oscillators provides an unprecedented control opportunity over the active phase that we can finally exploit.

\section*{Control of collective activity}
\begin{figure*}[h]
    \centering
    \includegraphics[width=\textwidth]{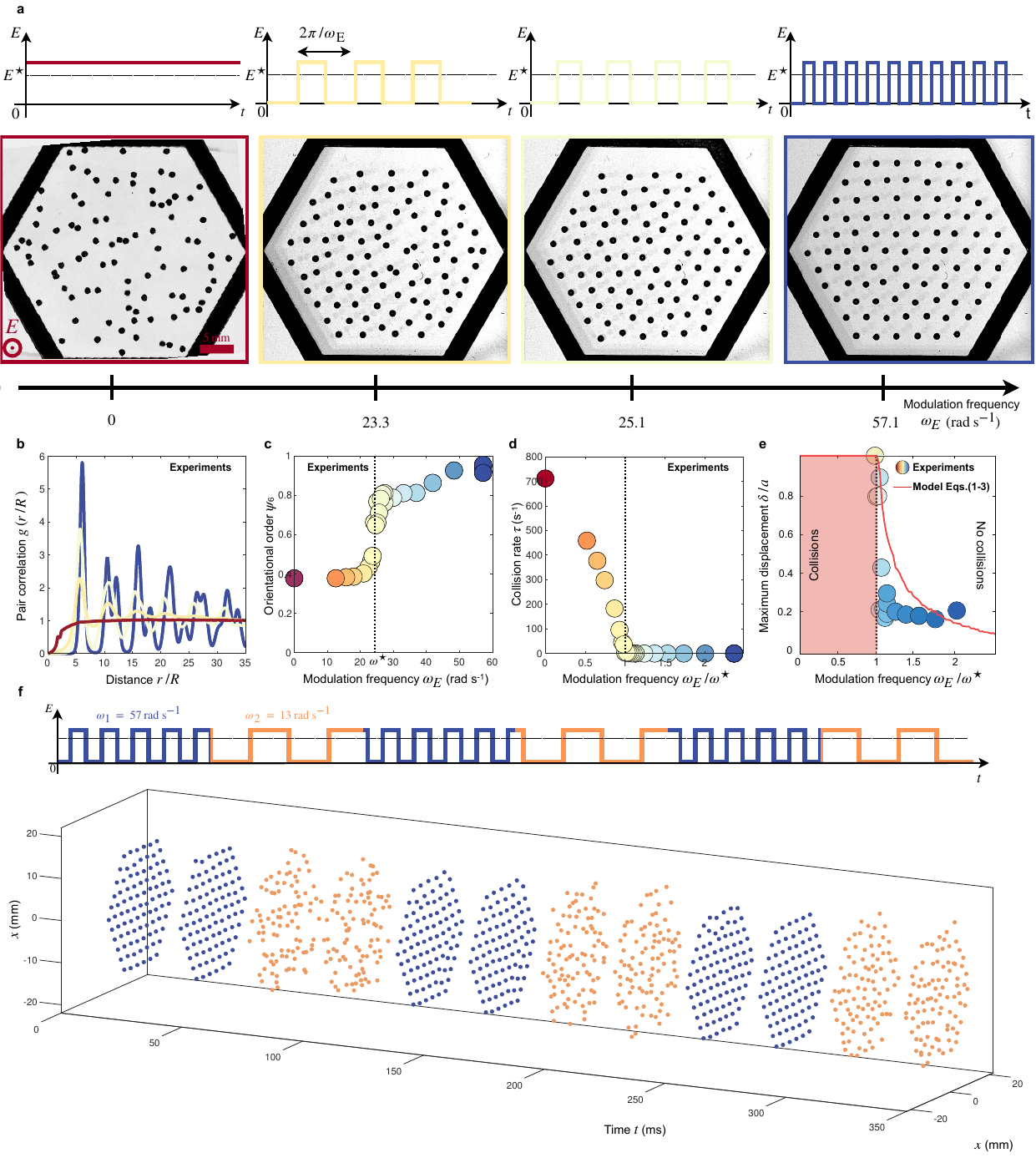}
    \caption{
{\bf Controlling collective activity to crystallize an oscillator gas. $\lvert$}
			{\bf (a)} Increase in ordering of the oscillator collective with the modulation frequency $\omega_E$. The highly disordered oscillator gas ($\omega_E = 0$) gives place to a more structured fluid ($\omega_E = 23.3\,\rm rad\,s^{-1}$), that subsequently crystallizes ($\omega_E \geq 25.1\,\rm rad\,s^{-1}$).
            Schematics of the electric signals applied are depicted on the top row: the square signal all have an amplitude $E_{\rm AC} = 0.87\,\rm kV\,mm^{-1} > E^{\star}$.
			{\bf (b)} Pair correlation functions for the four modulation frequencies of (a).
			{\bf (c)} The fluid-to-crystal transition is sharply captured by the gain of orientational order quantified by $\psi_6$. The transition is located at \textcolor{bleu}{$\omega^\star = 24\,\rm rad\,s^{-1}$}.
			{\bf (d)} The collision rate corroborates the control of collective activity: it monotonously decreases in the fluid phase until it vanishes at $\omega^\star$.
            {\bf (e)} Maximum displacement $\delta$ normalized by the lattice spacing $a$ \textcolor{bleu}{for the experiments (large dots) and the numerical simulations (red lines).} 
            \textcolor{bleu}{Numerical integration of Eqs.~(\ref{Eq:Phase}-\ref{Eq:Modulation}) for a pair of particles with initial spacing $a$ captures the transition from a state with collisions at low $\omega_E$ (shaded area) to a state with no collisions at high $\omega_E$.
            In this minimal two-body model, the transition occurs at \textcolor{bleu}{$\omega^\star_{\rm th} = 20.5\,\rm rad\,s^{-1}$}, in good agreement with the value found experimentally at the collective level (a-d).}
			{\bf (f)} The reversible transitions between active crystals (blue, $\omega_E = 57\,\rm rad\,s^{-1}$) and active gas (orange, $\omega_E = 13\,\rm rad\,s^{-1}$) illustrate the unique control offered by collective activity to channel the organization of active matter.
    }
    \label{Figure:4}
\end{figure*}
In order to control collective activity, attractive interactions between oscillators which eventually lead to super-elastic collisions must be limited.
The phase-dependence of Eq.~(\ref{Eq:Space}) gives a unique possibility to this regard.
Were synchrony, with $K_{ij}>0$, be favored over phase opposition, with $K_{ij}<0$, repulsive attractions would stabilize the system, as is the case in the so-called ``macroscopic Wigner crystals''~\cite{saint2004relaxation,coupier2006elasticite}.
Informed by the two-oscillator dynamics and its electric origin, we realize experimentally the modulation of $K_{ij}$ by applying an alternating current electric field $E_{\rm AC}$ with frequency $\omega_{E}$, in place of the DC field used so far.
\textcolor{bleu}{The dynamics of single oscillators under AC fields are reported in Extended Data Fig.~\ref{Figure:ED_SingleAC}.
At the collective level,} this strategy is successful: as $\omega_{E}$ increases, the oscillators self-organize into \textcolor{bleu}{states} of increasing order, eventually crystallizing on a triangular lattice \textcolor{bleu}{see Fig.~\ref{Figure:4}(a) and Supplementary Video~4}.
The increase in spatial correlations is captured by the pair-correlation functions $g(r)$, \textcolor{bleu}{(Fig.~\ref{Figure:4}(b))}.
From the very flat correlation of the active gas ($E=E_{\rm DC}$, {\it i.e.} $\omega_{E} = 0$), finite-range correlations emerges, before clear peaks revealing the symmetry of a hexagonal crystal rise.
Besides spatial correlations, orientational order also builds up.
Plotting the bond-orientation order parameter $\psi_6$ (see Methods) in Fig.~\ref{Figure:4}(c), we evidence a transition from a fluid phase to a crystal phase at \textcolor{bleu}{$\omega^\star \simeq 24\,\rm rad\,s^{-1}$.}
This transition \textcolor{bleu}{value} coincides with the visual inspection of the system, see Fig.~\ref{Figure:4}(a).

Examining the dynamics reveals that this structuring follows from the \textcolor{bleu}{successful control of collective activity as $\omega_{E}$ increases}. 
Indeed, as shown in Fig.~\ref{Figure:4}(d), the collision rate decreases with $\omega_{E}$ and vanishes continuously at \textcolor{bleu}{$\omega^\star$, reducing correspondingly the in-plane kinetic energy (see Extended Data Fig.~\ref{Figure:ED_4}).}
\textcolor{bleu}{In order to capture numerically the transition to a state with no collision, we augment the model of the binary dynamics Eqs.~(\ref{Eq:Phase}-\ref{Eq:Space}) to account for the time-modulation} of the synchronization coefficient $K_{ij}$.
We set:
\color{bleu}
\begin{align}
     K_{ij}(t) = K_0e^{-\Delta x_{ij}/\lambda}\cos(\omega_{E}t), \label{Eq:Modulation}
\end{align}

where the dependency on the inter-particle distance $\Delta x_{ij}$ is as before.

\color{black}
Integrating the equations of motion Eqs.~(\ref{Eq:Phase}-\ref{Eq:Modulation}) for pairs of oscillators, we recover the transition, as $\omega_{E}$ increases, between states with and without collisions\textcolor{bleu}{, and find $\omega^\star_{\rm th} = 20.5\,\rm rad\,s^{-1}$ (Fig.~\ref{Figure:4}(e))}.
This simple description validates our approach to limit collective activity \textcolor{bleu}{through the global tuning} of the phase dynamics.

\section*{Conclusion}
\textcolor{bleu}{We close this Article by stressing the unique control opportunity offered by the positive coupling between density and activity among oscillator collectives that} can be reversibly crystallized or melted in seconds, see Fig.~\ref{Figure:4}(f).
\color{bleu}
In the absence of thermal fluctuations or noise, the interaction-induced mobility allows the exploration of the configuration space, and the perfection of crystalline order.
We believe that our findings will motivate further investigations of this novel self-organization pathway.
In particular, while the active gas properties are not dependent on the system size (see Extended Data Fig.~\ref{Figure:ED_1} for a system of 919 oscillators), determining the nature of the crystallization transition and optimal annealing-like protocols require larger system-sizes that are hard to achieve experimentally.
Numerical and theoretical approaches would be highly beneficial to explore collective activity further and answer these challenging questions.

More generally, our findings provide new insights on the self-organization of active building blocks with responsive behaviour. 
They call in particular for a thorough exploration of active matter with positive density-activity coupling.
Going beyond the specificity of super-elastic collisions between non-motile building blocks, further investigations would establish and delineate universal properties of this new class of active materials.

\section*{Acknowledgements} 
We thank M. van der Veen, B. Durà Faulì, and C.M. Meulblok, for help with preliminary experiments.
We thank P.J. van Veldhuizen and R. Koehler from the Leiden Institute of Physics Electronics Department and R. Schrama and G. Verdoes from the Fine Mechanical Department for their help and expertise in developing the experimental setup.
We warmly thank D. Kraft and J. Palacci for helpful comments, and D.~Bartolo and M. van Hecke for insightful discussions.

\section*{Author contributions}
M. LB. and J. S. have equally contributed to this work.
A. M. and M. LB. designed the project. 
M. LB. and J. S. performed the experiments.
A. M. performed the finite element simulations. 
J. S. performed the numerical simulations.
All authors discussed the results and wrote the manuscript.

\section*{Competing interests}
The authors declare that they have no competing interests.

\section*{Data availability}
The raw datasets generated during the current study comprise several terabytes of data, it is not possible to share them on a public repository. However, they are available from the corresponding author on reasonable request. Source data for the figures are provided with this article~\cite{figshare}.

\twocolumn
\clearpage

\clearpage

\onecolumn
\clearpage
\section*{Methods}

\subsection*{Experimental setups}
We use two different designs for the millifluidic devices.

\textcolor{bleu}{Device $\#1$ - To perform the experiments pertaining to collective gas and its crystallization, we used a device in which the particles are allowed to move freely in the horizontal ($xy$)-plane while oscillating vertically (see Fig.~\ref{Figure:1}(a)).} 
The device is made of two glass slides coated with indium-tin oxide (ITO) acting as electrodes spaced by a \textcolor{bleu}{distance $h=3\,\rm mm$} \textcolor{bleu}{with a} PMMA (polymethyl methacrylate) frame.
The spheres are placed in a confining hexagon of side length $2\,\rm cm$ and height $2.8\,\rm mm$ made of metal and PMMA stacked on top of each other.
\textcolor{bleu}{
The metal layer is in contact with the bottom electrode and  repel the oscillators away from the boundary.
}

\textcolor{bleu}{Device $\#2$ - The experiments on the two-body dynamics are done with a device where we restrict the in-plane motion of the particles to one dimension, along $x$. 
The geometry is shown in Fig.~\ref{Figure:3}(a).}
The device is made of two metallic electrodes spaced by two PMMA spacers of height $h = 3\,\rm mm$.
The electrodes are $3\,\rm mm$-thick metal slabs placed in between two (non-conductive) glass slides.
\textcolor{bleu}{The particles are thereby confined to a region of $4\,\rm mm\times\, 3\,\rm mm\times\,1.8\,\rm mm$, and are imaged from the side in the ($xz$)-plane.}

\textcolor{bleu}{Materials - }The ITO-coated glass slides have dimensions $75\,\rm mm\,\times\,50\,\rm mm\,\times\, 1.1\, \rm mm$ and are coated with ITO (Solems, ITOSOL30).
They are connected to a high voltage amplifier (Matsusada, AMT-5B20-LC($230$V)) controlled by a function generator (Keysight, 33210A)\textcolor{bleu}{, which supplies either a direct current electric field $E_{\rm DC} = \lvert\mathbf{E}_{\rm DC}\lvert$ or an alternating current electric field with peak-to-peak amplitude $E_{\rm AC}$.}
In all experiments we use stainless steel spheres of diameter $2R = 1\rm\,mm$ and mass $m = 4\,\rm mg$ (Spherotech, AISI 420C) immersed in pure hexadecane (Sigma Aldrich).

\subsection*{Data acquisition}
\begin{figure*}[h]
    \centering
    \includegraphics[width=\textwidth]{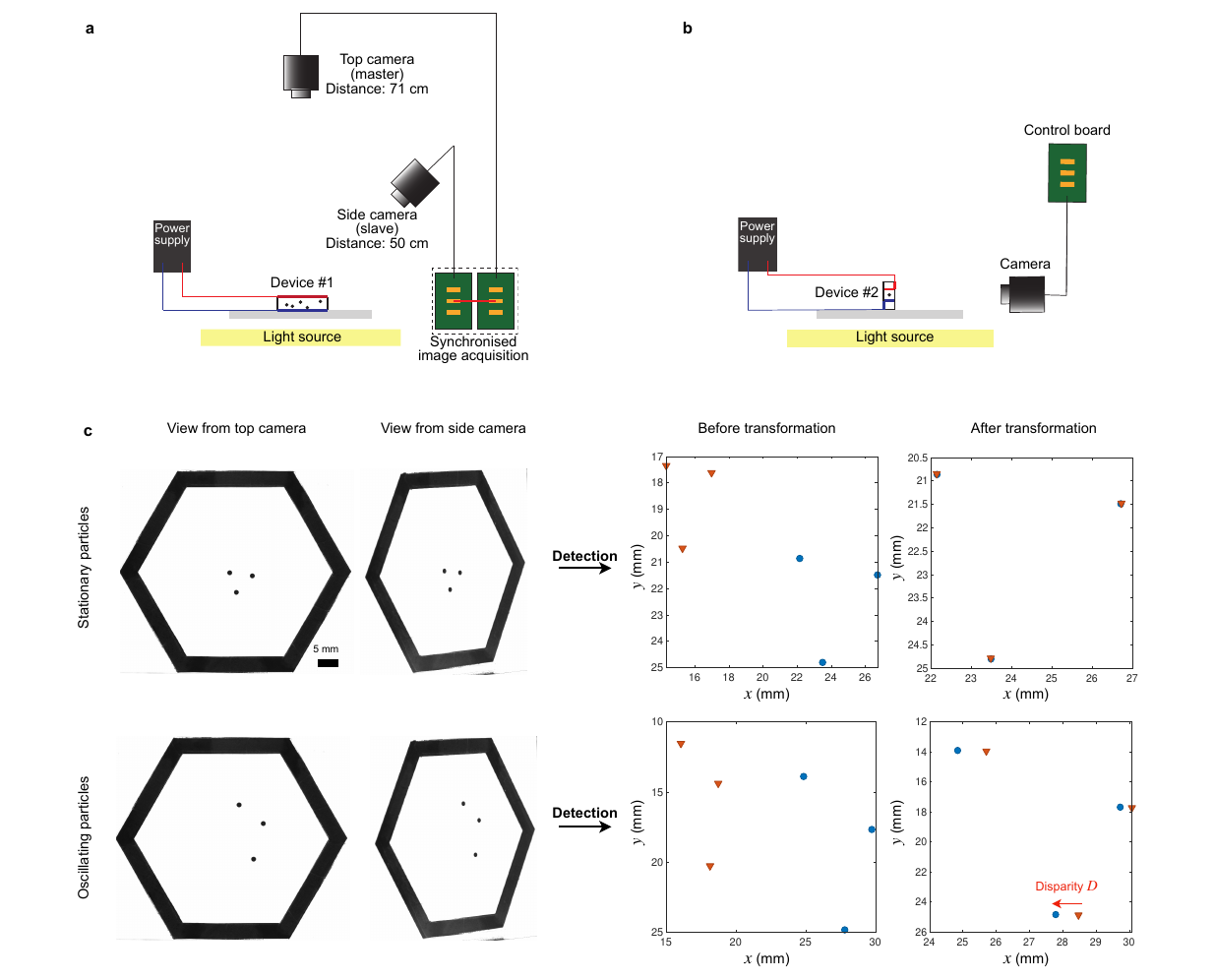}
    \caption{
{\bf Stereo-reconstruction of particle trajectories. $\lvert$}
			{\bf (a)} Camera configuration used for the stereo reconstruction of the \textcolor{bleu}{out-of-plane dynamics of the particles in the Device $\#1$.}
			{\bf (b)} Camera configuration used to image the \textcolor{bleu}{out-of-plane dynamics of the particles in the Device $\#2$} from the side.
            {\bf (c)} \textcolor{bleu}{Stereo-reconstruction in Device $\#1$. Top row: Stationary particles serve to calibrate the transformation and yield mathcing coordinates.
            Bottom row: The elevation of oscillating particles is proportional to de disparity $D$ between the reference coordinates and the transformed ones.
            Column 1: View from top camera.
            Column 2: Synchronous view from side camera.
            Column 3: Particle coordinates before transformation. Top view reference coordinates: blue dots. Side view coordinates: red triangles.
            Column 4: Particle coordinates after transformation.}
            The disparity $D$ is a measure for the height of the particles.
    }
    \label{Figure:SI_1}
\end{figure*}

\subsubsection*{Camera recording}
\textcolor{bleu}{Device $\#1$ -} To acquire the full 3-D dynamics of the particles, we use two cameras (Ximea, MX050MG-SY-X2G2-FL) to record the videos.
\textcolor{bleu}{The cameras are positioned as illustrated in Fig.~\ref{Figure:SI_1}(a).
One camera is placed directly above the device and image the ($xy$)-plane.
This top camera, equipped with an objective MVL100M23 and a tube lens CML15, is placed at a distance of $71\,\rm cm$ from the top electrode.
The second camera is oriented at a $45\,\rm deg$-angle from the $y$-axis.
This side camera, equipped with an objective MVL50M1 and a tube lens CML05, is placed at a distance of $50\,\rm cm$ from the top electrode.
The recordings are made on the two cameras with a framerate of $350\, \rm fps$.
The two cameras are synchronized in a master/slave configuration in order to reconstruct the dynamics of the particles using stereo-reconstruction.}

\textcolor{bleu}{Device $\#2$ - As illustrated in Fig.~\ref{Figure:SI_1}(b), a single camera, equipped with an objective MVL50M1 (and no tube lens), is placed to the side of the device. The framerate for acquisition ranges between $900-1000\,\rm fps$.}

\subsubsection*{Detection and tracking}
We use the MATLAB function \verb|imfindcircle| or the Fiji Analyze Particles module~\cite{schindelin2012fiji} to extract the positions of the particles.
The trajectories are then reconstructed using an adapted version of the MATLAB function by Blair and Dufresne based on the Crocker and Grier tracking algorithm~\cite{crocker1996methods,blair2008matlab}.

\subsubsection*{Stereo reconstruction}
\textcolor{bleu}{To obtain the full three-dimensional dynamics of the particles in Device $\#1$, we apply stereo reconstruction techniques to the synchronous acquisitions by the top and side cameras.}
\textcolor{bleu}{The top camera gives immediate access to the ($xy$)-motion.
The $z$-motion is obtained in two steps}.

\textbf{Step one: Calibration}\\
The first step of the reconstruction process is the calibration.
It consists in \textcolor{bleu}{determining the} transformation $H$ which convert the coordinates as seen from the side camera as if it was recorded from the \textcolor{bleu}{same plane as the top camera. 
The situation is then that of binocular vision (3D vision from two eyes).}
To find $H$ we first need \textcolor{bleu}{a pair of} images where we have located and \textcolor{bleu}{paired} two sets of corresponding points.
To this aim we take images of a black and white checkerboard from both the top and the side.
We subsequently locate the corners of the squares using the MATLAB function \verb|detectCheckerboardPoints|.
We use these sets of points as inputs of the OpenCV function \verb|cv2.findFundamentalMat| using an 8-point algorithm~\cite{longuet1981computer}, which gives us the \textcolor{bleu}{so-called} fundamental matrix $F$.
Using the function \verb|cv2.steroRectifyUncalibrated| and $F$, we find the homographies $H_1$ and $H_2$ corresponding to the top and side view respectively.
\textcolor{bleu}{
Next we normalize the homography $H_2$ in order to have the top view as the reference: $H = H_1^{-1}H_2 + \eta$.
At this stage, the constant $\eta$ is adjusted so that, applying $H$ to the sets of points (the checkerboard corners) acquired from the side camera yield transformed coordinates that perfectly match the top camera ones.}

\textcolor{bleu}{We next adjust the calibration in order to match the coordinates of particles {\em laying} on the bottom electrode inside the device, in place of the checkerboard corners.
To do so} we look for an additional affine transformation using the MATLAB function \verb|fitgeotrans|.
Using \textcolor{bleu}{the resulting} transformation, we are able to convert the \textcolor{bleu}{coordinates of the particles} obtained from the side camera as if recording from the top, as desired.

\textbf{Step two: Stereo-matching}\\
When the particles are laying on the bottom electrode, the transformed coordinates ($\tilde{x}_{\rm side},\tilde{y}_{\rm side})$ and references coordinates (${x}_{\rm top},{y}_{\rm top})$ match.
However, when the particles oscillate, the two sets of coordinates differ, as illustrated in Fig.~\ref{Figure:SI_1}(c-d).
Taking $\Delta x = x_{\rm top} - x_{\rm side}$ and $\Delta y = y_{\rm top} - y_{\rm side}$, we calculate the so-called disparity $D$ as
\begin{align}
    \textcolor{bleu}{D = sgn(\Delta x)\sqrt{\Delta x^2 + \Delta y^2}.}
\end{align}

Here the sign function is used to distinguish between shifts towards the left or towards the right.
The reconstructed height $z_{\rm reconstructed}$ of the particle is linearly related to $D$:
\begin{align}
    z_{\rm reconstructed} = \frac{z_{\rm max}}{D_{\rm max}}D,
\end{align}
where $D_{\rm max}$ is the maximum disparity measured during the recording and $z_{\rm max} = 2\, \rm mm$ is the maximum elevation of the (center of the) spheres.

\subsection*{Data analysis}
All calculations have been performed using the standard statistical toolkits in MATLAB.

\subsubsection*{\textcolor{bleu}{Definition of the oscillator phase $\theta$}}
\textcolor{bleu}{
We define the phase $\theta_i(t)$ of the particle $i$ from its instantaneous elevation $z_i(t)$ and velocity $\dot{z}_i(t)$ such that $\theta = \pi$ corresponds to the apex of the trajectory:}
\begin{align}
    \theta_i(t) = \begin{cases}
        \frac{\pi}{h-2R}\left( z_i(t)-\frac{h}{2} \right) + \frac{\pi}{2}, & \dot{z} \geq 0\\
        -\frac{\pi}{h-2R}\left( z_i(t)-\frac{h}{2} \right) + \frac{3\pi}{2}, & \dot{z} < 0,\\
    \end{cases}
\end{align}
\textcolor{bleu}{where $h=3\,\rm mm$ is the gap between the two electrodes and $2R=1\,\rm mm$ is the diameter of the particles.
In this definition, $z=0$ is taken at the bottom electrode and the top electrode is at $z=h$.}

\subsubsection*{\textcolor{bleu}{Kinetic energies}}
\textcolor{bleu}{The instantaneous velocitities are defined as $\mathbf{v}(t) = \left( \mathbf{r}(t)-\mathbf{r}(t-\Delta t)\right)/\Delta t$, with $\mathbf{r}(t) = (x(t),y(t),z(t))$ the position of the particle and $\Delta t = 1/{\rm fps}$, with ${\rm fps}$ the recording framerate.}
The \textcolor{bleu}{in-plane and out-of-plane energies respectively} are defined as
\textcolor{bleu}{\begin{align}
    \begin{split}
        &K_{\parallel} = \frac{m}{2}\left(v_x^2 + v_y^2 \right),\\
        &K_{\perp} = \frac{m}{2}v_z^2,
    \end{split}
\end{align}
}where $m = 4\,\rm mg$ is the mass of the oscillator.
The kinetic energies shown in Fig.~\ref{Figure:1} are averaged over the fast oscillatory motion using a moving average filter with a width of $8\,\rm frames = 23\,\rm ms$.
We note that Fig.~\ref{Figure:2}(a) shows the average kinetic energy within the collective $\langle K_{\perp,\parallel} \rangle$, while Fig.~\ref{Figure:1} reports energies of single oscillators.

\subsubsection*{Distributions}
The velocity distributions in Fig.~\ref{Figure:2} are constructed by only considering particles within a square region in the centre of the device with sides of length $17\,\rm mm$ in order to disregard boundary effects.

\subsubsection*{Collision analysis}
We define a collision event as a set of successive frames during which two particles have a distance less than $1.04\,\rm mm$.
We define $v_{\rm in}$ as a particle's velocity two frames before the first frame of the collision event and $v_{\rm out}$ is defined as its velocity two frames after the last frame of the collision event.

\subsubsection*{Orientational order}
We calculate the hexatic order parameter as
\begin{align*}
    \psi_6 = \frac{1}{\mathbf{card}(\mathcal{N}_i)}\sum_{j \in \mathcal{N}_i} e^{i6 \varphi_{ij}},
\end{align*}
where $\varphi_{ij} = \arctan\left(\frac{y_j-y_i}{x_j-x_i}\right)$ is the angle between the $x$-axis and the inter-particle vector joining particles $i$ and $j$.
The elements of the set $\mathcal{N}_i$ that contain the six nearest neighbours of particle $i$ are calculated using a Voronoi tessellation.

\subsection*{Finite-element simulations}
We compute the electric potential energy $E$ for pairs of oscillators by finite element simulations.
We use the electrostatic module of the software COMSOL Multiphysics 5.5.
The two spheres of radius $R=0.5\,\rm mm$ are centred in a large square domain of linear size $20\,\rm mm \gg R$, caped by two electrodes spaced by $3\,\rm mm$.
The two electrodes are grounded, so only the interaction energy contributes to the electric energy.
The pre-defined settings for the mesh are used, with ``finer'' element sizes.

We simulate the system for all combinations of particle separations $\delta x = x_{\rm right}-x_{\rm left} - 2R$ between $0\,\rm mm$ to $3\,\rm mm$ by steps of $0.25\,\rm mm$, and particle elevations ranging from the bottom to the top electrodes by steps of $0.1\,\rm mm$.
The latter corresponds to phase differences between 0 and $2\pi$ by steps of $\pi/10$. 
Each oscillator bears a charge $\lvert q \lvert = 4\pi\epsilon_r\epsilon_0 R^2 E_{\rm DC} \simeq 9.5\times 10^{-11}\,\rm C$, whose sign depends on its phase. 
In this expression, $\epsilon_0$ is the vacuum permittivity, $\epsilon_r = 2.05$ is the dielectric constant of hexadecane, and $E_{\rm DC} = 1.7\,\rm kV\,mm^{-1}$.
Simulations with both positive charges, both negative charges, or opposite charges are made.

 The average electric potential energy $\langle E \rangle_{T_{\rm o}}$ is calculated as the average over one oscillation of the potential energy $E$. 
 Practically, it is obtained by consolidating the relevant charge combinations, for some given particle separation and phase-difference, along one oscillation.
The resulting two-dimensional energy map is shown in the main text Fig.~\ref{Figure:3}(c).

\subsection*{Numerical integration of the equations of motions}
\label{Methods:Finite-element}
We numerically integrated Eqs.~(\ref{Eq:Phase}-\ref{Eq:Modulation}) using Heun's method with a timestep of $\Delta t = 1\times 10^{-5}\,\rm s$.
The coupling terms $K_{ij}(t)$ and $J_{ij}$ are taken
\begin{align}
    &K_{ij}(\Delta x_{ij}, t) = K_0\exp(-\Delta x_{ij}/\lambda)\cos(\omega_{\rm AC}t),\\
    &J_{ij}(\Delta x_{ij}) = J_0\exp(-\Delta x_{ij}/\lambda),
\end{align}
with $K_0 = -3.927\,\rm rad\,s^{-2}$, $J_0=-3.105\,\rm m\,s^{-2}$ and $\lambda = 8.052\times 10^{-3}\,\rm m$ calculated from the finite-element simulations.
The oscillation period is taken $T_{\rm 0} = 0.01\, \rm s$, in agreement with the experiments.
\textcolor{bleu}{The relaxation timescales $T_\theta = T_x = 0.02\,\rm s$} are assumed to be equal and have been calculated from $T_{\rm o}$ and the restitution coefficient $e\sim 0.8$ (see SI).
To drive the \textcolor{bleu}{synchronization}, we add a small white-noise term to Eq.~(\ref{Eq:Phase}) with a strength of $1\times 10^{-4}$ $\rm rad\,s^{-2}$.
We have checked that increasing this value by one order of magnitude does not significantly change the values reported in Fig.~\ref{Figure:4}(e).
We simulate the dynamics of two particles with an initial separation $a$ of $4\,\rm mm$. Due to the periodicity of $K_{ij}(t)$, the relative displacements $\delta$ of the particles reaches a limit-cycle where the width of the cycle depends on the modulation frequency $\omega_{E}$. Once $\delta/a=1$, the particles collide and the solid melts. We find that this happens for a critical modulation frequency \textcolor{bleu}{$\omega^\star = 20.5\,\rm rad\,s^{-1}$.}

\newpage

\onecolumn
\section*{Methods-only references}

\clearpage

\onecolumn
\newpage
\renewcommand\thesection{}
\begin{center}
\section*{Supplementary Information}
\end{center}

\cftsetindents{section}{0em}{3em}
\cftsetindents{subsection}{2em}{3.5em}
\cftsetindents{subsubsection}{4em}{4.5em}
\renewcommand{\contentsname}{}
\pagenumbering{gobble}
\tableofcontents
\thispagestyle{empty}
\pagenumbering{arabic}
\newpage

\setcounter{figure}{0}
\renewcommand\thesection{SI.\arabic{section}}
\setcounter{equation}{0}

\section{Self-oscillations obtained through contact-charge electrophoresis}
\label{Section:SI_CCEP}
\renewcommand{\thefigure}{S\arabic{figure}}
\renewcommand{\theequation}{S\arabic{equation}}

\begin{figure*}[h!]
	\includegraphics[width=\textwidth]{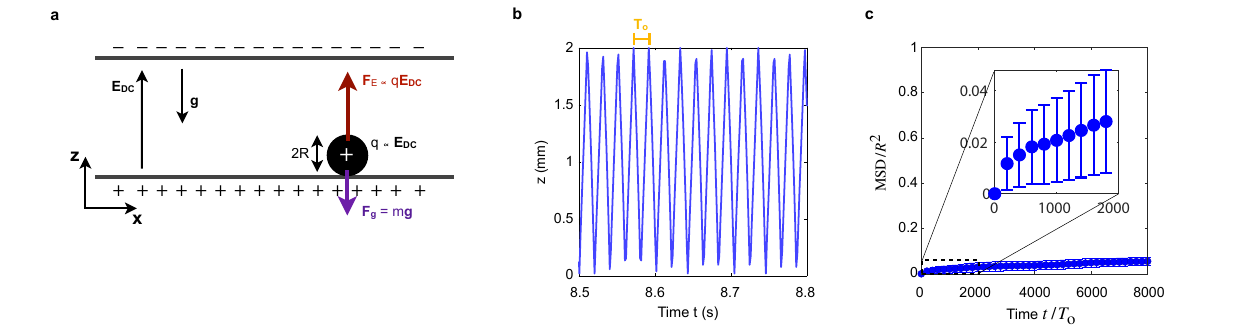}
	\caption{{\bf Contact-charge electrophoresis: dynamics of an isolated oscillator. $\lvert$}
            \textcolor{bleu}{{\bf (a)} Sketch of the forces at play during CCEP. An electrically conductive particle of mass $m$ and radius $R$ resting on the bottom, positive, electrode acquires a charge of same polarity. Purple arrow: weight. Red arrow: electric force.}
			{\bf (b)} Oscillatory motion $z(t)$ of the particle along the direction of the applied electric field $E = 0.87\,\rm kV\,mm^{-1}$.
			{\bf (c)} The mean-squared displacement (MSD) scaled by the square of the particle radius $R$ over a time of $8000$ periods of oscillation \textcolor{bleu}{i.e. about $2\,\rm min\,30\,s$}. The oscillator barely moves. \textcolor{bleu}{The errorbars show one standard deviation of uncertainty. Inset: Zoom-in of the MSD at short times.}
		}

    \label{Figure:SI_3}

\end{figure*}
Contact-charge electrophoresis (CCEP) drives the oscillatory motion of conductive particles between two electrodes using a constant electric field (see Fig.~\ref{Figure:SI_3}(a)).
The mechanism~\cite{mersch2011antiphase,drews2015contact} is as follow:
A particle of mass $m$ initially resting on the bottom electrode acquires a charge proportional to its area and to the electric field $q \propto R^2 E_{\rm DC}$ by virtue of Gauss' law.
It experiences thereby an electric force $F_{E} \propto q E_{\rm DC} \propto R^2 E_{\rm DC}^2$ which attracts it towards the top electrode.
Only when this electric force overcomes gravity $F_{E} > mg$ (where $g = 9.81\,\rm m\,s^{-2}$ is the gravitational acceleration constant), does the particle lift off.
By studying a single particle in the Device $\#2$, we determine the electric field threshold to be $E_{\rm exp}^{\star} = 0.72\,\rm kV\,mm^{-1}$, which agrees well with the theoretical value $E_{\rm th}^{\star} = 0.69\,\rm kV\,mm^{-1}$.

Once the particle reaches the top electrode, it acquires the reverse charge and gets electrically attracted towards the bottom electrode.
Oscillatory motion results from the repetition of this process.

As shown in Extended Data Fig.~\ref{Figure:ExtData_SingleOscilator}(a-c), a steady-state is quickly reached, characterized by nearly triangular trajectories as shown in Fig.~\ref{Figure:SI_3}(b) and a well-defined oscillation period $T_{\rm o}$.
As discussed in the main text and shown in Extended Data Fig.~\ref{Figure:ExtData_SingleOscilator}(d), the oscillation frequency is measured to increase linearly with the electric field amplitude $\omega_{\rm o}\sim E_{\rm DC}$.
This dependency is not only advantageous to the study of collective activity - since it amounts to a tuneable energy reserve of the active gas building blocks - it is also indicative of the dominant dissipation mechanism.
Indeed, were drag with the solvent the main dissipation mechanism, the scaling would be $\omega_{\rm o}\sim E_{\rm DC}^2$.
Instead, the linear scaling reveals that inelastic collisions with the electrodes dominate dissipation:
they amount to a drag force proportional to $\omega_{\rm o}^2$ which balances the electric force in $E_{\rm DC}^2$ (see also Section~\ref{Sec:Dissipation}).

Finally, we show in Fig.~\ref{Figure:SI_3}(e), the in-plane mean-squared displacement of an isolated oscillator in the \textcolor{bleu}{Device $\#1$} over 8000 periods of oscillation ($\simeq 160\,\rm s$):
its dynamics is extremely localized.
This illustrates that individual oscillators are \textcolor{bleu}{not only non-motile, but their in-plane dynamics is also} virtually noiseless.

\section{Rescaling of velocity distributions}
\label{Section:SI_velocity}
The collapse of the velocity distributions shown in Fig.~\ref{Figure:2}(e) is explained by the scaling with $\omega_{\rm o}$ shared between the energy transferred from the oscillatory, \textcolor{bleu}{out-of-plane}, motion to the \textcolor{bleu}{in-plane} degrees of freedom on one hand and the dissipation of \textcolor{bleu}{in-plane} kinetic energy on the other hand.

\subsection{ Scaling of energy injection.}
\begin{figure*}[h!]
	\includegraphics[width=\textwidth]{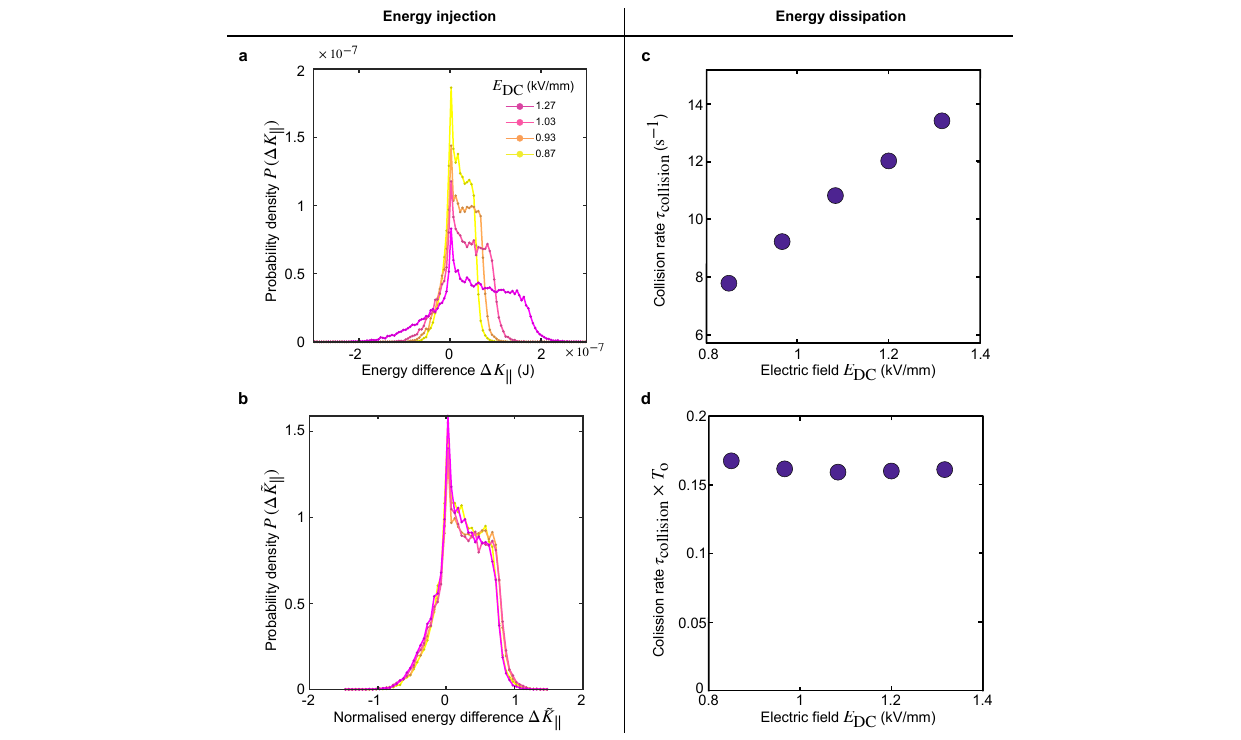}
	\caption{{\bf Active gas: rescaling of velocities. $\lvert$}
			{\bf (a-b)} Probability distributions of the variation of \textcolor{bleu}{in-plane kinetic energy $\Delta K_\parallel = m(v^2_{\rm out} - v^2_{\rm in})/2$} at collisions\textcolor{bleu}{, with $v_{\rm in}$ and $v_{\rm out}$ the particle speed before and after collision respectively (see Methods)}. 
            The distributions broaden with \textcolor{bleu}{$E_{\rm DC}$} and collapse upon normalizing \textcolor{bleu}{$\Delta K_\parallel$} with the \textcolor{bleu}{out-of-plane kinetic energy $\widetilde{K}_\parallel = K_\parallel/(h\omega_{\rm o}/\pi)$. The color codes for the out-of-plane kinetic energy are the same as in Fig.~\ref{Figure:2}(a).}
			{\bf (c-d)} The collision rate $\tau_{\rm collisions}$ increases linearly with $E$ and remains constant upon normalization by $1/T_{\rm o}$.
		}
            \label{Figure:SI_4}

\end{figure*}

As shown in the joint probability density $P(v_{\rm in}, v_{\rm out})$ in Fig.~\ref{Figure:2}(d), $v_{\rm out}$ takes values up to the oscillation speed $v_{\rm o}\sim 200\,\rm mm\,s^{-1}$. 
This suggests that increasing the energy of oscillation by increasing \textcolor{bleu}{the electric field amplitude} $E_{\rm DC}$ should similarly increase the energy transferred to the $(x,y)$-plane \textcolor{bleu}{via super-elastic binary} collisions. This is confirmed by measuring the change in the \textcolor{bleu}{particle in-plane kinetic energy} $\Delta K_\parallel = m(v^2_{\rm out} - v^2_{\rm in})/2$ at collisions \textcolor{bleu}{(where $v_{\rm in}$ is the particle speed just before a collision and $v_{\rm out}$ its speed just after, as in the Main text).
Figure~\ref{Figure:SI_4}(a) shows that the distributions of $\Delta K_\parallel$ broaden as $E_{\rm DC}$ is increased. 
Upon rescaling the energies by the oscillation energy $\Delta \Tilde{K}_\parallel = \Delta K_\parallel/(h\omega_{\rm o}/\pi)$ the distributions collapse, see Fig.~\ref{Figure:SI_4}(b).}

\subsection{Scaling of energy dissipation.} \label{Sec:Dissipation}
We now turn to the dissipation occurring along the trajectories after \textcolor{bleu}{binary} collisions.
\color{bleu}
Every time a single oscillator hits an electrode, their inelastic collision reduces the particle speed by a factor $e_\parallel$ (a so-called restitution coefficient), such that after $n$ such collisions, the particle speed is given by:
\begin{align}
v(n) = v_{\rm out}e_\parallel^n = v_{\rm out} \exp(n \ln{e_\parallel}).\label{Eq:Inelastic}
\end{align}
Since a collision with an electrode occurs every half oscillation period, we change the variable $n$ in Eq.~(\ref{Eq:Inelastic}) for $t = n T_{\rm o}/2$:
\begin{align}
v(t) = v_{\rm out} \exp\left(-\frac{t}{T_{\rm o}/(2\ln{(1/e_\parallel))}}\right) = v_{\rm out}\exp(-t \tau_{\rm dissipation}), \label{Eq:dissipation}
\end{align}
where we introduced the dissipation relaxation rate $\tau_{\rm dissipation}$.
Equation~\eqref{Eq:dissipation} shows that inelastic collisions with the electrodes result in an effective Stokes drag occurring on a timescale set by the oscillation frequency $\omega_{\rm o}$:
\color{black}
\begin{align}
    \tau_{\rm dissipation} = \frac{\omega_{\rm o}}{\pi}\ln\left(\frac{1}{e_\parallel}\right)\propto \omega_{\rm o}. \label{Eq:DissipInelastic}
\end{align}

\subsection{Scaling of collision rate.}
\color{bleu}
After a binary collision, the particle speed typically decays as
\begin{align}
v(t) = v_{\rm out}\exp(-t \tau_{\rm dissipation}),\;t\in[0,\,t_{\rm collision}], \label{Eq:Speed}
\end{align}
where $t_{\rm collision}$ is the typical time between binary collisions defined by equating the mean-free-path $l$ with the distance travelled:
\begin{align}
    l = \frac{v_{\rm out}}{\tau_{\rm dissipation}}\left[ 1-\exp\left( -t_{\rm collision}\tau_{\rm dissipation} \right) \right]. \label{Eq:meanFreePath}
\end{align}
Since $l$ is a geometrical quantity (given by the system density and the particle radius), it is independent of $\omega_{\rm o}$.
Solving Eq.~\eqref{Eq:meanFreePath} for $t_{\rm collision}$, we get
\begin{align}
    t_{\rm collision} = -\frac{\ln\left({1-\frac{l\tau_{\rm dissipation}}{v_{\rm out }}}\right)}{\tau_{\rm dissipation}}\propto \frac{1}{\omega_{\rm o}},
\end{align}
where the last proportionality follows from the previous paragraphs. 
We confirm this scaling experimentally, and show in Fig.~\ref{Figure:SI_4}(c-d) that the system collision rate is proportional to $\omega_{\rm o}$.

\subsection{Rescaling of the velocity distributions.}
The rescaling of the velocity distributions can now be easily rationalized.
Indeed, rescaling the equation of motion Eq.~\eqref{Eq:Speed} with $t' = t \omega_{\rm o}$ and $v' = v \omega_{\rm o}$ yields:
\begin{align}
v'(t')= \frac{v_{\rm out}}{\omega_{\rm o}} \exp{(-t' \tau_{\rm dissipation}/\omega_{\rm o})}, \;t'\in[0,\,t_{\rm collision}\omega_{\rm o}].
\end{align}
Since $v_{\rm out} \propto \omega_{\rm o}$, $\tau_{\rm dissipation} \propto \omega_{\rm o}$, and, $t_{\rm collision} \propto 1/\omega_{\rm o}$, this dynamics is independent of $\omega_{\rm o}$.
As a consequence, the rescaled velocity distributions are also independent of $\omega_{\rm o}$, in agreement with our experimental findings (Main text Fig.~\ref{Figure:2}(e)).
\color{black}
Altogether our findings validate the description of the active gas as being dominated and fuelled by super-elastic collisions.

\color{bleu}
\textcolor{bleu}{
\section{Derivation of the equations of motion with phase-dependent interactions}
We derive the equations of motion Eqs.~\eqref{Eq:Phase}-\eqref{Eq:Space}, Main text, from first principles starting from the general Newton equation.
We first show that the periodic nature of the oscillations in the $z$-coordinate leads to a natural representation of the equations in terms of their phase $\theta\in[0,2\pi]$.
Upon coarse-graining the resulting equations over one oscillation period $T_{\rm o}$, the interaction forces take on a simple form in terms of a Fourier cosine series.
We next obtain the values of the strength of interactions and discuss timescales of dissipation.
Finally, we stress that hydrodynamics effects do not play a significant role.}

\subsection{Periodic wrapping of $z$-coordinate}
The general Newton equation for an oscillator reads
\begin{align}
    \label{Eq:Newton}
    m\ddot{\mathbf{r}}_i = \mathbf{F}_{\rm drive} + \mathbf{F}_{\rm drag} + \mathbf{F}_{\rm interaction}
\end{align}
where $\mathbf{r_i}(t) = \left(x_i(t),z_i(t)\right)$ is the oscillator position, $m$ its mass, $\dot{\square} \equiv \partial\square/\partial t$ denotes time derivative, and the forces are defined by:
\begin{align}
    \label{Eq:Force_definition}
    \begin{split}
        \mathbf{F_{\rm drive}}(\dot{z}_i)& \equiv \mathrm{sgn}\left(\dot{z}_i\right)\lvert \mathbf{F}_{\rm drive}\lvert \mathbf{\hat{z}},\\
        \mathbf{F_{\rm drag}}(\dot{x}_i, \dot{z}_i)&\equiv -\left( \gamma_\parallel \dot{x}_i \mathbf{\hat{x}} + \gamma_\perp\dot{z}_i\mathbf{\hat{z}} \right),\\
        \mathbf{F_{\rm interaction}}(x_i, z_i, \dot{x}_i, \dot{z}_i,x_j, z_j, \dot{x}_j, \dot{z}_j) &\equiv f_{\rm polarity}(\dot{z}_i, \dot{z}_j)f_{\rm magnitude}(\Delta x_{ij}, z_i, z_j)\mathbf{\hat{r}_{ij}}.
    \end{split}
\end{align}
The driving term $\mathbf{F}_{\rm drive}$ is the electric attraction to the oppositely-charge electrode underlying contact-charge electrophoresis.
The drag term $\mathbf{F}_{\rm drag}$ accounts for dissipation with in-plane and out-plane magnitudes $\gamma_\parallel$ and $\gamma_\perp$, respectively.
The interaction force $\mathbf{F}_{\rm interaction}$ between particles $i$ and $j$ has been factorized into two contributions: a polarity contribution $f_{\rm polarity}(\dot{z}_i, \dot{z}_j):=\mathrm{sgn}\left(\dot{z}_i\dot{z}_j\right)$ to keep track of the relative signs of the charges of the particles and a magnitude $f_{\rm magnitude}\geq0$.
We stress that the magnitude depends on both $z_i$ and $z_j$ rather than $\Delta z_{ij}$ due to the presence of the two electrodes.
Finally, $\mathbf{\hat{r}_{ij}}$ is the unit vector joining particle j to i, such that its projections onto $\hat{x}$ and $\hat{z}$ read:
\begin{align}
\label{Eq:Projection_coefficients}
    \begin{split}
        P_x&=\mathbf{\hat{r}}_{ij}\cdot\hat{x} = \frac{\Delta x_{ij}}{\sqrt{\Delta x_{ij}^2 + \Delta z_{ij}^2}},\\
        P_z&=\mathbf{\hat{r}}_{ij}\cdot\hat{z} = \frac{\Delta z_{ij}}{\sqrt{\Delta x_{ij}^2 + \Delta z_{ij}^2}},
    \end{split}
\end{align}
where $\Delta x_{ij} = x_j-x_i$ and $\Delta z_{ij} = z_j-z_i$.

Using these definitions, the projection of Eq.~\eqref{Eq:Newton} onto $\mathbf{\hat{x}}$ and $\mathbf{\hat{z}}$ yields the set of coupled equations
\begin{align}
\label{Eq:Newton_XZ}
\begin{split}
    m\ddot{x}_i &= -\gamma_\parallel \dot{x}_i + f_{\rm polarity}f_{\rm magnitude}P_x,\\
    m\ddot{z}_i &= \mathrm{sgn}\left( \dot{z}_i \right)\lvert F_{\rm drive}\lvert-\gamma_\perp \dot{z}_i + f_{\rm polarity}f_{\rm magnitude}P_z.    
\end{split}
\end{align}

We next invert the definition of $\theta$ (Methods):
\begin{align}
\label{Eq:Height_to_Phase}
    \begin{cases}
        z_i(\theta_i) = \frac{h}{2} + \frac{2}{\pi}\left( \frac{h}{2}-R \right) \left(\theta_i-\frac{\pi}{2}\right),\;\dot{z}_i>0\\
        z_i(\theta_i) = \frac{h}{2} + \frac{2}{\pi}\left( \frac{h}{2}-R \right) \left(\frac{3\pi}{2}-\theta_i\right),\;\dot{z}_i<0
    \end{cases}
\end{align}
where $h$ is the height difference between the two electrodes and $R$ is the radius of the particles.
In addition, the $n$-th order derivatives of $z_i$ and $\theta_i$ for $n\geq 1$ are related via
\begin{align}
    \frac{\partial^n\theta_i}{\partial t^n}=\frac{\pi}{h-2R}\mathrm{sgn}\left( \dot{z}_i \right)\frac{\partial^n z_i}{\partial t^n}. \label{Eq:Derivatives}
\end{align}

Using the relations Eqs.~\eqref{Eq:Height_to_Phase} and~\eqref{Eq:Derivatives}, the $z$-equation~\eqref{Eq:Newton_XZ} can be written as
\begin{align}
    \begin{split}
        \mathrm{sgn}\left( \dot{z}_i \right)\frac{\partial^2\theta_i}{\partial t^2}=&\frac{1}{m}\frac{\pi}{h-2R}\mathrm{sgn}\left( \dot{z}_i \right)\lvert F_{\rm drive}\lvert - \frac{\gamma_\perp}{m}\mathrm{sgn}\left( \dot{z}_i \right)\frac{\partial \theta_i}{\partial t}\\
        &+\frac{1}{m}\frac{\pi}{h-2R}\mathrm{sgn}\left( \dot{z}_i \right)\mathrm{sgn}\left( \dot{z}_j \right) f'_{\rm magnitude} P'_z, \label{Eq:signed}
    \end{split}
\end{align}
where we have used the definition of $f_{\rm polarity}$ and introduced prime notations to denote $f_{\rm magnitude}$ and $P$ expressed as a function of the variables $\theta_i$ and $\theta_j$.
Importantly, Eq.~\eqref{Eq:signed} can be simplified by $\mathrm{sgn}({\dot{z}_i})$, such that it does not depend on it anymore: when the dynamics is represented in terms of phase, the ascending and descending parts of the trajectories are described by the same equation:
\begin{align}
\label{Eq:Phase_intermediate}
         \frac{\partial^2 \theta_i}{\partial t^2}&=\frac{\omega_{\rm o}}{T_{\theta}}-\frac{1}{T_{\theta}}\frac{\partial \theta_i}{\partial t}+\frac{1}{m}F_{\rm interaction, \theta},
\end{align}
where $T_{\theta}=\frac{m}{\gamma_\perp}$ is the out-of-plane relaxation time, $\omega_{\rm o} = \frac{{T_\theta}}{m}\frac{\pi}{h-2R}\lvert F_{\rm drive}\lvert$ is the oscillation frequency of individual particles, and
\begin{align}
        F_{\mathrm{interaction},\theta}(\Delta x_{ij}, \theta_i, \theta_j) =\frac{\pi}{h-2R} \mathrm{sgn}({\dot{z}_j})f'_{\rm magnitude}(\Delta x_{ij}, \theta_i, \theta_j)P_z'(\Delta x_{ij},\theta_i,\theta_j),
\end{align}
is the interaction term.
We stress that $F_{\mathrm{interaction},\theta}$ is only a function of $\Delta x_{ij}$, $\theta_i$, and $\theta_j$, since $\mathrm{sgn}(\mathbf{\dot{z}_j}) = \mathrm{sgn}(\theta_j-\pi)$.

A similar treatment for the $x$-projection yields the equation of motion
\begin{align}
\label{Eq:Space_intermediate}
         \frac{\partial^2 x_i}{\partial t^2}&=-\frac{1}{T_x}\frac{\partial x_i}{\partial t}+\frac{1}{m}F_{\mathrm{interaction}, x},
\end{align}
with $T_x = \frac{m}{\gamma_{\parallel}}$, and
\begin{align}
        F_{\mathrm{interaction},x}(\Delta x_{ij}, \theta_i, \theta_j) =f'_{\rm polarity}(\theta_i, \theta_j)f'_{\rm magnitude}(\Delta x_{ij}, \theta_i, \theta_j)P_x'.
\end{align}

\subsection{Coarse-graining over time}
We next simplify the expressions for the interaction terms by averaging the dynamics over one oscillation period $T_{\rm o}$.
We do so by assuming, based on our experimental observations, that $\Delta \theta_{ij}$ is approximately constant over a period $T_{\rm o}$.
The averaged interaction potential $\langle{V}_{\rm interaction}\rangle_{T_{\rm o}}$ can therefore be written as a function of $\Delta\theta_{ij}$ rather than a function of both $\theta_i$ and $\theta_j$.
It can further be expressed as a Fourier cosine expansion due to the fact that the potential is periodic in $\Delta\theta_{ij}$ with a period of $2\pi$ and that it's an even function in $\Delta\theta_{ij}$ due to the symmetry of the system geometry around $z=h/2$:
\begin{align}
\label{Eq:Fourier_Cosine}
    \langle{V}_{\rm interaction}(\Delta x_{ij}, \Delta \theta_{ij})\rangle_{T_{\rm o}} = \sum_{n\geq 1}\widetilde{K}_n(\Delta x_{ij})\cos(n\Delta\theta_{ij}).
\end{align}
Retaining the lowest frequency component gives
\begin{align}
    \langle{V}_{\rm interaction}(\Delta x_{ij}, \Delta \theta_{ij})\rangle_{T_{\rm o}} = \widetilde{K}(\Delta x_{ij})\cos(\Delta\theta_{ij}).
\end{align}
The time averaged force components can now be calculated by taking the appropriate derivatives.
Taking the average over $T_{\rm o}$ of Eqs.~\eqref{Eq:Phase_intermediate} and~\eqref{Eq:Space_intermediate} and omitting the averaging brackets $\langle\dots\rangle$ for clarity yields the coupled equations (Main text Eqs.~\eqref{Eq:Phase}-~\eqref{Eq:Space}):
\begin{align}
    \begin{split}
        \frac{\partial^2 x_i}{\partial t^2}&=-\frac{1}{T_x}\frac{\partial x_i}{\partial t}+J_{ij}\cos(\Delta\theta_{ij}),\\
         \frac{\partial^2 \theta_i}{\partial t^2}&=\frac{\omega_{\rm o}}{T_{\theta}}-\frac{1}{T_{\theta}}\frac{\partial \theta_i}{\partial t}+K_{ij}\sin(\Delta\theta_{ij}),
    \end{split}
\end{align}
where the interaction coefficients are given by
\begin{align}
&K_{ij}(\Delta x_{ij}) =-\frac{1}{m}\frac{\pi}{h-2R}\widetilde{K}(\Delta x_{ij}), \label{Eq:CoeffK} \\
&J_{ij}(\Delta x_{ij}) =-\frac{1}{m}\frac{\partial \widetilde{K}(\Delta x_{ij})}{\partial x_i}. \label{Eq:CoeffJ}
\end{align}

\subsection{Determination of the interaction coefficients and relaxation timescales}

\subsubsection{Interaction coefficients}

\begin{figure*}[h!]
	\includegraphics[width=\textwidth]{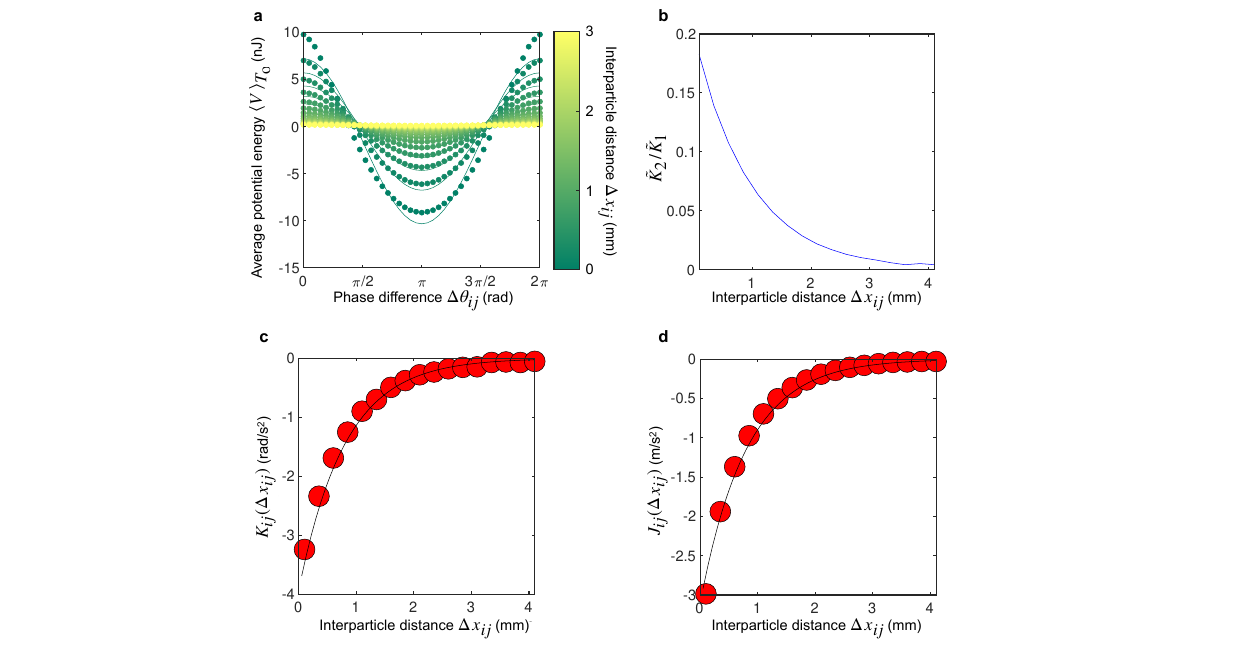}
	\caption{\textcolor{bleu}{{\bf Fourier cosine expansion of the averaged potential energy. $\lvert$}
            \textbf{(a) } Fourier cosine expansion (Eq.~\eqref{Eq:Fourier_Cosine}) up to $1$-st order (solid lines) of the averaged potential energy calculated by finite-element simulations (dots).
            \textbf{(b) } Ratio $\widetilde{K}_2/\widetilde{K}1$ of the amplitudes of the first two harmonics $n$ in the Fourier cosine expansion. The averaged potential energy is well approximated already by the lowest frequency contribution for low inter-particle distances $\Delta x_{ij}$, but this approximation improves even more for larger inter-particle distances.
            \textbf{(c)} Variation of the phase interaction coefficient $K_{ij}$ as a function of the inter-particle distance $\Delta x_{ij}$. Red dots: full force obtained from finite-element simulations. Solid line: force obtained via Eq.~\ref{Eq:CoeffK} and the exponential fit of $\widetilde{K}_{ij}$.
            \textbf{(d)} Variation of the in-plane interaction coefficient $J_{ij}$ as a function of the inter-particle distance $\Delta x_{ij}$. Red dots: full force obtained from finite-element simulations. Solid line: force obtained via Eq.~\ref{Eq:CoeffJ} and the exponential fit of $\widetilde{K}_{ij}$.}
		}
    \label{Figure:fourierAmplitudes}

\end{figure*}

First, we confirm that the averaged potential energy is indeed well approximated by keeping only the lowest frequency of its Fourier expansion.
Figure~\ref{Figure:fourierAmplitudes}(a) shows cuts of $\langle V \rangle_{T_{\rm o}}$ (obtained by finite element simulations and shown in Main text Fig.~\ref{Figure:3}(c)) for different values of $\Delta x_{ij}$.
The first harmonics closely approximate the interaction energy for all interparticle distances.
Additionally, we compare in Fig.~\ref{Figure:fourierAmplitudes}(b) the first-two harmonics $\widetilde{K}_1$ and $\widetilde{K}_2$: even at contact, $\widetilde{K}_2/\widetilde{K}_1 < 20\%$.

Second, we extract $\widetilde{K}$ from the simulation and find that it decays exponentially as $\widetilde{K}(\Delta x_{ij}) = \widetilde{K}_0\exp{(-\Delta x_{ij}/\lambda)}$, with $\widetilde{K}_0 = 1\times10^{-8}\,\rm J$ and $\lambda = 8.052\times10^{-4}\,\rm m$.
This exponential decay of electric interaction is generally expected from the confinement between two electrodes, see e.g.~\cite{galatola2006determination}.
We show in Fig.~\ref{Figure:fourierAmplitudes}(c-d) the resulting $K_{ij}(\Delta x_{ij})$ and $J_{ij}(\Delta x_{ij})$ obtained through Eqs.~\eqref{Eq:CoeffK}-\eqref{Eq:CoeffJ}, which compares very well with the (computed) full electric couplings.

Finally, we obtain
\begin{align}
    &K_{ij}(\Delta x_{ij}) = K_0 \exp\left( -\Delta x_{ij}/\lambda \right), \\
    &J_{ij}(\Delta x_{ij}) = J_0 \exp\left( -\Delta x_{ij}/\lambda \right),
\end{align}
with $K_0=-3.927\,\rm rad\,s^{-2}$ and $J_0=-3.105\,\rm m\,s^{-2}$.

\subsubsection{Relaxation timescales}

\begin{figure*}[h!]
	\includegraphics[width=\textwidth]{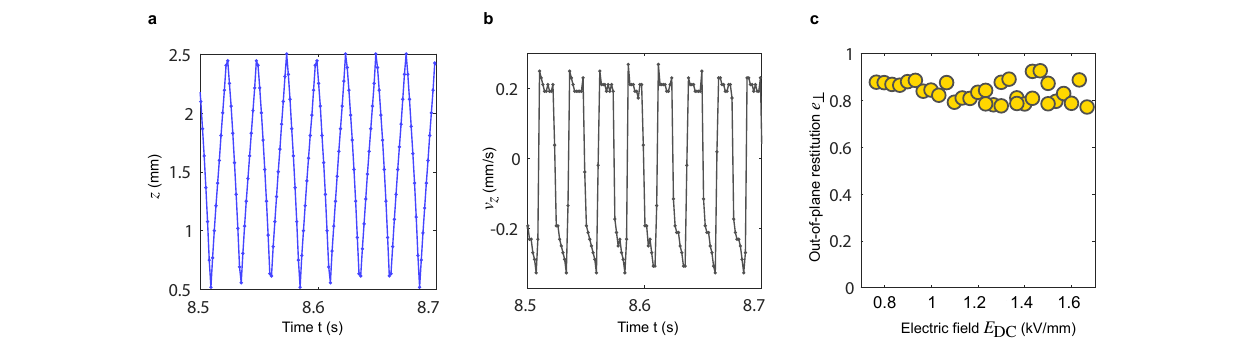}
	\caption{\textcolor{bleu}{{\bf Oscillatory trajectory and restitution coefficient $\lvert$}
            \textbf{(a)}~Oscillatory motion of an isolated oscillator over $200$ milliseconds. $E_{\rm DC} = 0.78\,\rm kV\,mm^{-1}$.
            \textbf{(b)}~Instantaneous velocity corresponding to the trajectory displayed in ($\mathbf{a}$). We can directly calculate the vertical restitution coefficient $e_\perp$ with the ratio of the speeds directly after and before contact with an electrode.
            \textbf{(c)}~We measure on an isolated oscillator and over $1500$ contacts with the electrodes the average out-of-plane restitution coefficient $e_\perp$ for $33$ values of the amplitude of the electric field $E_{\rm DC}$.}
		}
    \label{Figure:SM_restitution}
\end{figure*}

The linear scaling of the oscillation frequency with the electric field amplitude suggests that inelastic collisions between the oscillators and the electrodes is the dominant dissipation mechanism~\cite{mersch2011antiphase} (see Fig.~\ref{Figure:ExtData_SingleOscilator}(d)).
Consistently, we estimate the relaxation timescales $T_x$ and $T_\theta$ through Eq.~\eqref{Eq:DissipInelastic}.
As shown in Fig.~\ref{Figure:SM_restitution}, we measure a restitution coefficient of $e_\perp\simeq 0.8$.
Assuming the same value for $e_\parallel$, we take
\begin{align}
    &T_x = T_\theta = 0.02\,\rm s. \label{Eq:Tdissip}
\end{align}
Overall, the binary dynamics and the active gas-to-crystal transition are quantitatively captured by electric interactions and dissipation at electrodes.

\subsection{Neglecting hydrodynamic effects}
As a final note, all our findings suggest that hydrodynamic effects do not play a significant role.
Indeed, viscous dissipation occurs on a typical time-scale $T_\eta \sim 0.1\,\rm s$ (assuming Stoke's drag) which is much longer than the dissipation time-scale associated with inelastic collision at the electrodes Eq.~\eqref{Eq:Tdissip}.
Besides, neglecting hydrodynamic interactions in the equations of motion yields qualitative and quantitative agreement with the experimental observations.
This is confirmed by a rough estimate of hydrodynamic interactions given by Faxén's law $F_{\rm hydro}\simeq \eta d^2 v_{\rm o}/\Delta x_{ij} \sim 10^{-7}\,\rm N$, which is one order of magnitude lower than electric interactions of order $10^{-6\,\rm} N$ for particle separation $\Delta x_{ij} \sim 1\,\rm mm$.

\color{black}
\section{Supplementary videos}
\begin{itemize}
    \item{\bf Supplementary Video 1: Collective activity.} This video demonstrates the concept behind collective activity. An active gas emerges from otherwise noiseless oscillators at high densities. Density $\phi = 9\%$. Side video of internal dynamics slowed down by a factor $25$.\\
    \item{\bf Supplementary Video 2: Active oscillator gas.} This video illustrates the athermal nature of the oscillator gas: i) the coexistence between stationary (blue) and moving (green) particles, and ii) the breaking of time-reversal symmetry. Density $\phi = 9\%$. Videos slowed down by a factor $16$.\\
    \textcolor{bleu}{\item{\bf Supplementary Video 3: Binary collisions at the onset of the active gas.} This video illustrates that binary collisions occurs at the onset of the active gas. Density $\phi = 9\%$. Video slowed down by a factor of $16$.}\\
    \item{\bf Supplementary Video 4: Reversible melting transition.} This video illustrates how the control of collective activity leads to the structuring of the oscillator gas and its eventual crystallization. Density $\phi = 9\%$.\\
\end{itemize}

\newpage
\pagenumbering{gobble}
\pagenumbering{arabic}
\begin{center}
\section*{Extended Data}
\end{center}
\setcounter{figure}{0}
\renewcommand{\thefigure}{ED\arabic{figure}}

\begin{figure*}[h!]
	\includegraphics[width=\textwidth]{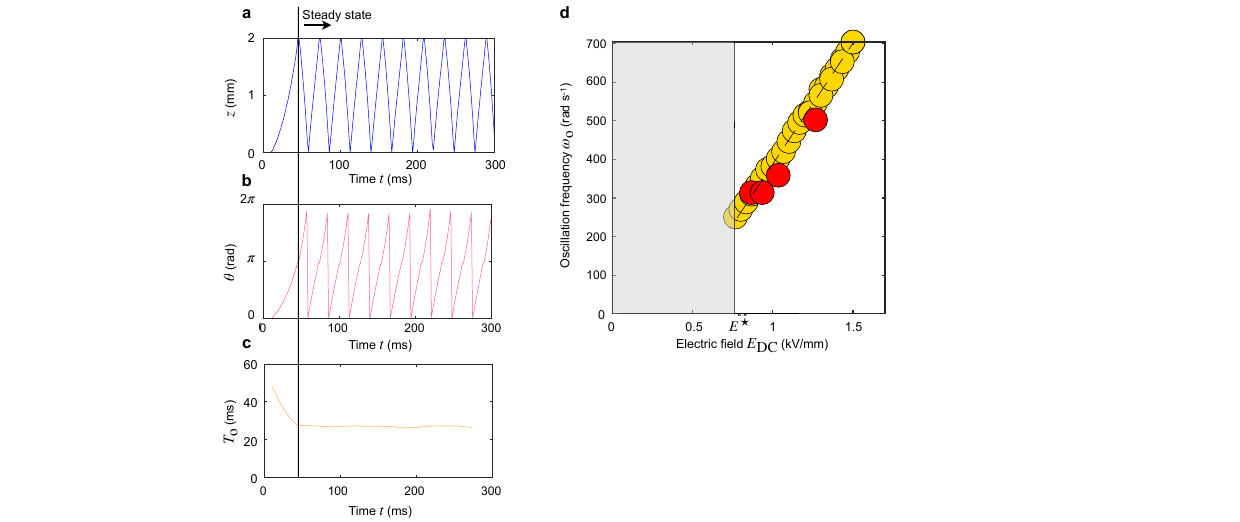}
	\caption{\textcolor{bleu}{{\bf Individual oscillator under constant electric field. $\lvert$}
			{\bf (a-c)} The steady-state of self-sustained oscillation is quickly reached after the electric field ($E_{\rm DC} = 0.77\,\rm kV\,mm^{-1}$) is applied ($t=0$).
            {\bf (a)} The particle elevation as a function of time is nearly triangular.
            {\bf (b)} Correspondingly, the particle phase is nearly linear in the interval $[0,\,2\pi[$.
            {\bf (c)} The instantaneous oscillation period $T(t)= t(\theta + 2\pi) - t(\theta)$ decays to the steady-state value $T_{\rm o}$ in $\sim 50\,\rm ms$.
            {\bf (d)} The period of oscillation $T_{\rm o}$ of a single oscillator is (yellow dots) is determined by the electric field amplitude $E_{\rm DC}$ as $1/T_{\rm o} \propto E_{\rm DC}$ (black line). 
            This scaling is consistent with dissipation dominated by inelastic collisions with the electrodes rather than fluid drag.
            Oscillations occur above a threshold value for the electric field $E^\star = 0.72\,\rm kV\,mm^{-1}$ materialized by the grey area.
            The red dots correspond to the average period of oscillations measured in the collective with packing fraction $\phi = 9\,\rm \%$.
		}}
	\label{Figure:ExtData_SingleOscilator}
\end{figure*}

\newpage
\begin{figure*}[h!]
	\makebox[\textwidth][c]{\includegraphics[width=\textwidth]{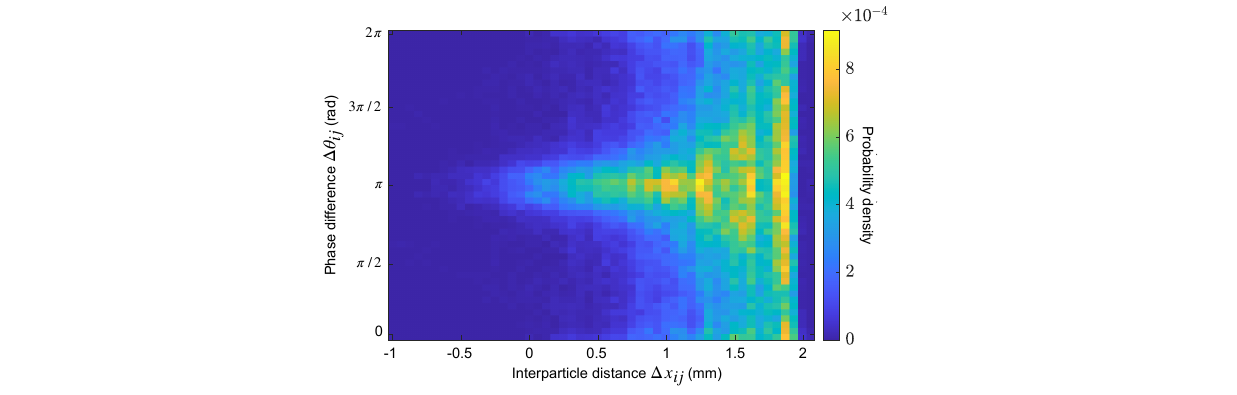}}
	\caption{\textcolor{bleu}{{\bf Dynamics of binary collisions. $\lvert$}
			The joint probability density of $(\Delta x_{ij},\,\Delta\theta_{ij})$ reveals that initially steady particles typically synchronize to $\Delta\theta_{ij} = \pi$ and then attract.
            Over 200 collisions between initially static particles are analysed. As in the main text, $\Delta x_{ij} = x_{j}-x_{i} - 2R$ and $\Delta \theta_{ij} = \theta_{\rm j} - \theta_{\rm i}$.
            The experimental path of highest probability is drawn on the Main-text Fig.~\ref{Figure:3}(c).
		}}
	\label{Figure:SI_2}
\end{figure*}

\newpage
\begin{figure*}[h!]
	\includegraphics[width=\textwidth]{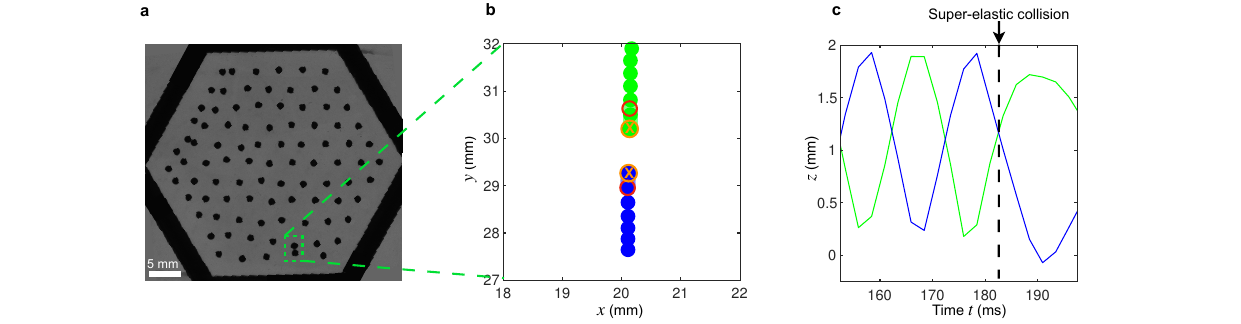}
	\caption{\textcolor{bleu}{{\bf Binary collisions at the onset of the active gas. $\lvert$}
            {\bf (a)} Snapshot of the system corresponding to the first binary collision (highlighted in green) after application of the electric field.
            {\bf (b)} Close-up showing the trajectories of the two particles colliding in (a). 
            Red circles indicate their initial positions. Orange circles and crosses indicate their positions at contact. 
            After their super-elastic collision, the two particles move quickly away from each others.
            {\bf (c)} Elevation of the two particles as a function of time reconstructed from stereo-imaging ($t=0$ when the field is applied). 
            Before they collide, the two particles oscillate in phase opposition.
            }
		}
	\label{Figure:ED_Onset}
\end{figure*}

\newpage
\begin{figure*}[h!]
	\includegraphics[width=\textwidth]{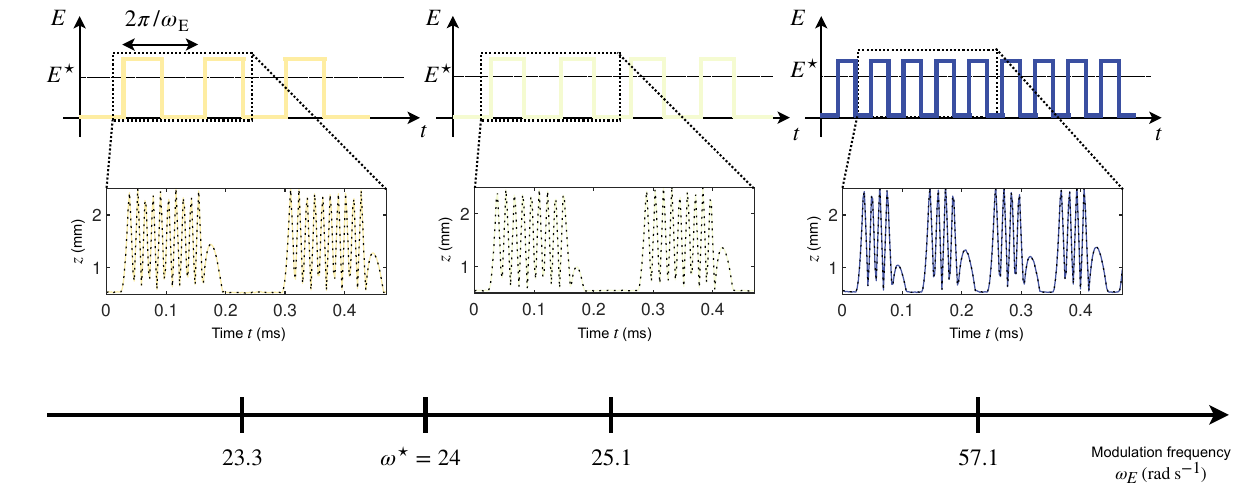}
	\caption{\textcolor{bleu}{{\bf Individual oscillator under alternative electric field. $\lvert$}
            Top row: schematics of the electric field signals with varying modulation frequency $\omega_{E}$.
            Bottom row: experimental trajectories of individual oscillators under these conditions.
            Oscillator heights $z$ under a square wave with increasing modulation frequency $\omega_{\rm E}$. 
            The values of $\omega_{E}$ (and corresponding colours) correspond to the data on the oscillator collective reported in the Main-text Fig.~\ref{Figure:4}(a). The dotted black line aids visibility.
            Importantly, the active gas-crystal transition occurs at $\omega^\star$ where the individual dynamics does not change qualitatively: the transition is a true collective phenomenon. $E_{\rm AC} = 1\,\rm{kV\,mm^{-1}}$.}
		}
	\label{Figure:ED_SingleAC}
\end{figure*}

\newpage
\begin{figure*}[h!]
	\includegraphics[width=\textwidth]{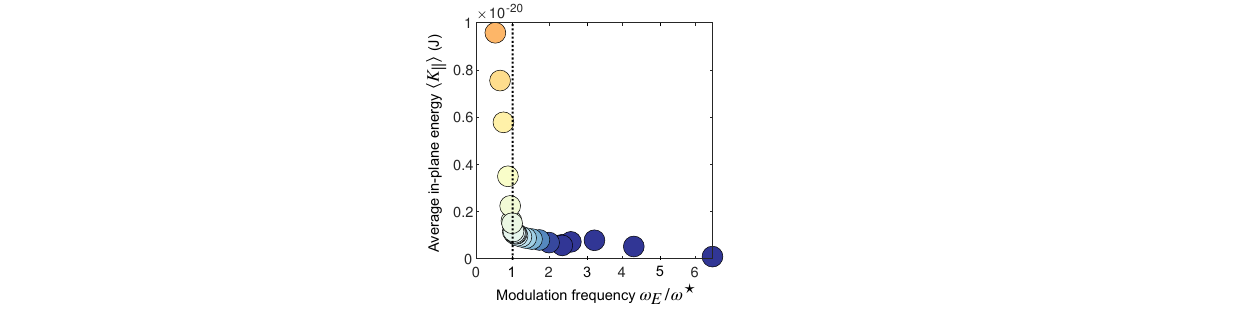}
	\caption{\textcolor{bleu}{{\bf Decrease in kinetic energy across the active gas-crystal transition. $\lvert$}
            The in-plane kinetic energy decreases sharply with the modulation frequency $\omega_{E}$ until $\omega^\star$.
            This decrease is concomitant with the decrease in the collision rate reported in the Main-text Fig.~\ref{Figure:4}(d): limiting the number of super-elastic collision is effective in taming collective activity.
		}
        }
	\label{Figure:ED_4}
\end{figure*}

\newpage
\begin{figure*}[h!]
	\includegraphics[width=\textwidth]{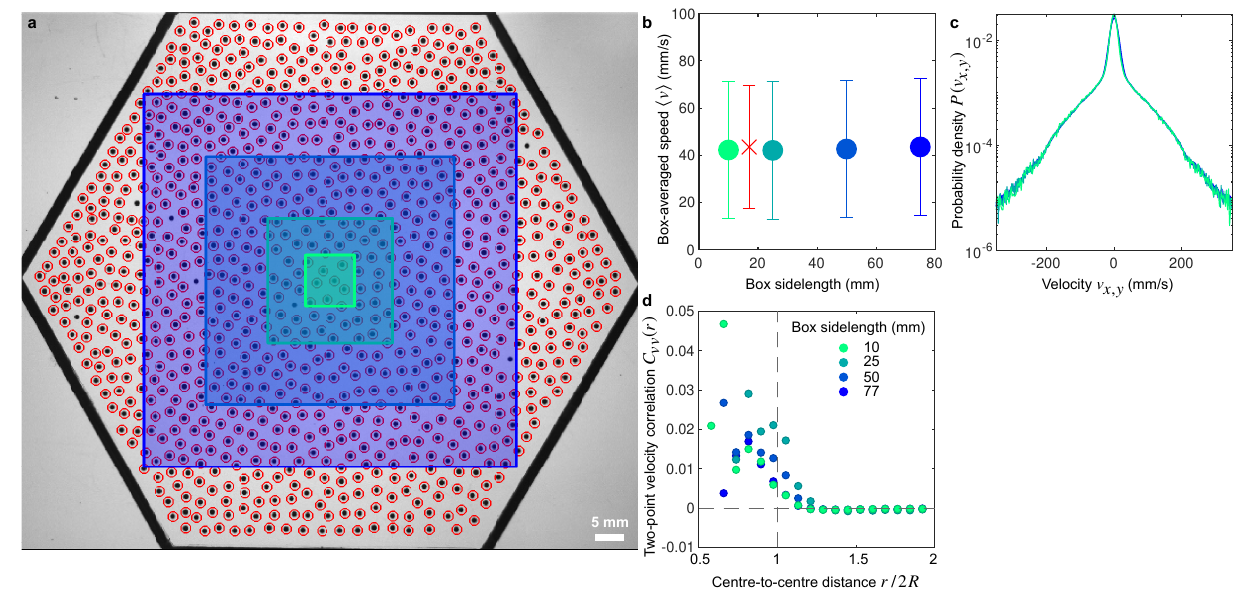}
	\caption{\textcolor{bleu}{{\bf Finite size analysis of the active gas. $\lvert$}
            {\bf (a)} Experimental snapshot of an active gas composed of 919 particles at a density $\phi = 9\%$. Red circles indicate the detected particles. Squares materialize the different box sizes used for the analysis of finite-size effects in panels (b-d).
            We use the same colour code from the smallest box (green) to the largest box (deep blue).
            {\bf (b)} Mean speeds $\langle v\rangle$ for different box sizes as a function of the side-lengths of the box. 
            No significant difference is found among the different box sizes (coloured dots), nor with the smaller system used in the main text (red cross). The errorbars show one standard deviation of uncertainty. The statistics have been derived from $N_{\rm samples} = \sum_{i=1}^{N_{\rm frames}}N_{\rm{particles, }i}$ samples, where $N_{\rm frames}$ is the total number of frames and $N_{\rm{particles}, i}$ is the total number of particles that have a well-defined velocity within the different box sizes at each frame $i$.
            For the system composed of 919 particles, $N_{\rm frames}=348729$. The total samples sizes for increasing box sizes are $N_{\rm samples}=169644,\,1074496,\,4274413,\,9555308,$ respectively. The total number of frames for the system composed of 91 particles is $N_{\rm frames}=95183$. Its total samples size is $N_{\rm samples}=8661653$.
            {\bf (c)} Probability density functions $P(v_{x,y})$ of the velocity components $v_x$ and $v_y$ for the different box sizes.
            The probability density functions are independent on the size of the box, suggesting that finite-size effects are negligible.
            {\bf (d)} The two-point velocity correlation $C_{vv}(r)$ for different box sizes.
            There are virtually no velocity correlations at distances greater than one particle diameter $r/2R = 1$.
            This extremely short correlation length validates that the active gas statistics reported in the Main text are not influenced by finite-size effects.}
		}
	\label{Figure:ED_1}
\end{figure*}

\end{document}